\documentclass[
 reprint,
superscriptaddress,
 amsmath,amssymb,
 aps,
prl,
floatfix,
english
]{revtex4-2}

\usepackage[subpreambles=true]{standalone}
\usepackage{import}
\usepackage{bibunits}
\defaultbibliography{references}
\defaultbibliographystyle{apsrev4-2}

\usepackage{graphicx}
\usepackage{svg}
\usepackage{dcolumn}
\usepackage{bm}
\usepackage{siunitx}
\usepackage[T1]{fontenc}
\usepackage[utf8]{inputenc}

\DeclareUnicodeCharacter{03C0}{$\pi$}
\usepackage[unicode=true, colorlinks=true, citecolor={blue!80!black}, urlcolor={blue!50!black}, linkcolor = {blue!80!black}]{hyperref}

\newcommand{\D}{\Delta}
\newcommand{\GL}{\Gamma_L}
\newcommand{\GR}{\Gamma_R}
\newcommand{\eo}{\epsilon_0}

\newcommand{\dfn}{\delta f_{r,n}}
\newcommand{\fbare}{f_{r,b}}
\newcommand{\Vgc}{V_{g}}

\newcommand{\cqed}{cQED}
\mathchardef\mhyphen="2D

\graphicspath{{Figures/}}

\begin{document}

\widetext

\title{
Microwave susceptibility observation of interacting many-body Andreev states
}

\author{V.~Fatemi}
\email{valla.fatemi@yale.edu}
\affiliation{Departments of Physics and Applied Physics, Yale University, New Haven, CT 06520, USA}
\author{P.~D.~Kurilovich}
\affiliation{Departments of Physics and Applied Physics, Yale University, New Haven, CT 06520, USA}
\author{M.~Hays}
\affiliation{Departments of Physics and Applied Physics, Yale University, New Haven, CT 06520, USA}
\author{D.~Bouman}
\affiliation{QuTech and Delft University of Technology, 2600 GA Delft, The Netherlands}
\affiliation{Kavli Institute of Nanoscience, Delft University of Technology, 2600 GA Delft, The Netherlands}
\author{T.~Connolly}
\affiliation{Departments of Physics and Applied Physics, Yale University, New Haven, CT 06520, USA}
\author{S.~Diamond}
\affiliation{Departments of Physics and Applied Physics, Yale University, New Haven, CT 06520, USA}
\author{N.~E.~Frattini}
\affiliation{Departments of Physics and Applied Physics, Yale University, New Haven, CT 06520, USA}
\author{V.~D.~Kurilovich}
\affiliation{Departments of Physics and Applied Physics, Yale University, New Haven, CT 06520, USA}
\author{P.~Krogstrup}
\affiliation{Center for Quantum Devices and Station Q Copenhagen, Niels Bohr Institute, University of Copenhagen, Universitetsparken 5, 2100 Copenhagen, Denmark}
\author{J.~Nygård}
\affiliation{Center for Quantum Devices, Niels Bohr Institute, University of Copenhagen, Universitetsparken 5, 2100 Copenhagen, Denmark}
\author{A.~Geresdi}
\affiliation{QuTech and Delft University of Technology, 2600 GA Delft, The Netherlands}
\affiliation{Kavli Institute of Nanoscience, Delft University of Technology, 2600 GA Delft, The Netherlands}
\affiliation{Quantum Device Physics Laboratory, Department of Microtechnology and Nanoscience, Chalmers University of Technology, SE 41296 Gothenburg, Sweden}
\author{L.~I.~Glazman}
\affiliation{Departments of Physics and Applied Physics, Yale University, New Haven, CT 06520, USA}
\author{M.~H.~Devoret}
\email{michel.devoret@yale.edu}
\affiliation{Departments of Physics and Applied Physics, Yale University, New Haven, CT 06520, USA}

\begin{abstract}

Electrostatic charging affects the many-body spectrum of Andreev states, yet its influence on their microwave properties has not been elucidated.
We developed a circuit quantum electrodynamics probe that, in addition to transition spectroscopy, measures the microwave susceptibility of different states of a semiconductor nanowire weak link with a single dominant (spin-degenerate) Andreev level.
We found that the microwave susceptibility does not exhibit a particle-hole symmetry, which we qualitatively explain as an influence of Coulomb interaction.
Moreover, our state-selective measurement reveals a large, $\pi$-phase shifted contribution to the response common to all many-body states which can be interpreted as arising from a phase-dependent continuum in the superconducting density of states.

\end{abstract}

\maketitle

Andreev states are the supercurrent-carrying fermionic modes that govern the electrodynamical response of Josephson devices.
In devices hosting only a few transport channels, individual states may become energetically well-separated and thus addressable.
This direct access to Andreev states has been leveraged to discover new phenomena in mesoscopic superconductivity and unveil applications such as Andreev qubits~\cite{bretheau_localized_2013,janvier_coherent_2015,van_woerkom_microwave_2017,hays_direct_2018,tosi_spin-orbit_2019,hays_continuous_2020,hays_coherent_2021}. 
These experiments probed the microwave frequency response of discrete Andreev states in different regimes: from a minimal configuration of an atomic point contact with one strongly dispersing Andreev level, to multi-state configurations in long nanowire weak links where spin-orbit effects become important.

Our understanding of the microwave frequency electrodynamics involving Andreev states has so far relied on noninteracting pictures focusing on the sub-gap levels~\cite{kos_frequency-dependent_2013,park_andreev_2017,tosi_spin-orbit_2019,park_adiabatic_2020}.
While such pictures describe atomic point contacts well~\cite{bretheau_localized_2013,janvier_coherent_2015}, they have two blind spots for any finite-length weak link.
First, charging energy should be present when the electrons experience a nonzero dwell time in the weak link.
Yet, to our knowledge, the impact of charging energy on the microwave response of Andreev states has not been investigated.
This is in contrast to measurements that have revealed a rich interplay between superconductivity and charging effects in dc transport through quantum dots~\cite{buitelaar_multiple_2003,van_dam_supercurrent_2006,jorgensen_critical_2007,pillet_andreev_2010,li_0-pi_2017,kim_transport_2013,lee_spin-resolved_2014,delagrange_manipulating_2015,szombati_josephson_2016,lee_scaling_2017,razmadze_quantum_2020,su_erasing_2020}.
Second, under the same conditions, the spectral continuum outside the superconducting gap (which we will refer to as "the continuum" for brevity) also contributes to the supercurrent~\cite{kulik_macroscopic_1970,beenakker_resonant_1992,meden_andersonjosephson_2019} and therefore is electrodynamically active in finite-length weak links. 
However, the dynamics of the continuum have not been isolated from the contributions of the subgap Andreev states.
This requires measurements that resolve individual many-body configurations of the system.

Are charging effects and the continuum relevant for the microwave response of superconductor-semiconductor weak links?
Here, we answer affirmatively by performing state-resolved microwave response measurements with a circuit quantum electrodynamics (\cqed) probe.
Application of \cqed~techniques has revealed different quasiparticle occupation configurations of Andreev levels~\cite{janvier_coherent_2015,hays_direct_2018,hays_continuous_2020,hays_coherent_2021}, which is possible because the different states of the quantum system have different electrodynamical susceptibility~\cite{blais_circuit_2021}.
However, beyond state determination, the magnitude and dispersion of the response functions of individual states carry a wealth of physical information~\cite{paila_current-phase_2009,basov_electrodynamics_2011,park_adiabatic_2020,metzger_circuit-qed_2021,haller_phase-dependent_2021}.
This information complements that of microwave and tunneling spectroscopy which are restricted to transitions between states of the same and opposite parity, respectively.

We measured a Josephson semiconductor nanowire with a single dominant Andreev level coupled to a superconducting microwave resonator.
We extracted the admittance and transition spectrum of different many-body quasiparticle configurations and find two qualitative differences to the standard theory~\cite{ivanov_two-level_1999,zazunov_andreev_2003,zazunov_dynamics_2005,kos_frequency-dependent_2013,beenakker_resonant_1992,wendin_josephson_1996}. 
First, the microwave responses of the even-parity states are not symmetric about the odd state, which we refer to as a violation of a particle-hole symmetry.
Second, all states exhibit a common $\pi$-shifted contribution to their phase dispersion.
We interpret these observations as the qualitative effects of a charging energy in the weak link and a phase-dispersing continuum, respectively.
Our recently developed theory~\cite{kurilovich_microwave_2021} can account for these discrepancies.
Finally, we measured the fermion parity polarization as a function of phase and gate voltage. 
The switches of the polarization are consistent with 0-$\pi$ transitions, while the incompleteness of the polarization indicates nonequilibrium parity dynamics. 
These observations lay a foundation for investigating the interplay of Coulomb interactions with superconducting pairing in conventional and topological superconducting mesoscopic devices with \cqed~probes.

\begin{figure}
\includegraphics[width=1.0\columnwidth]{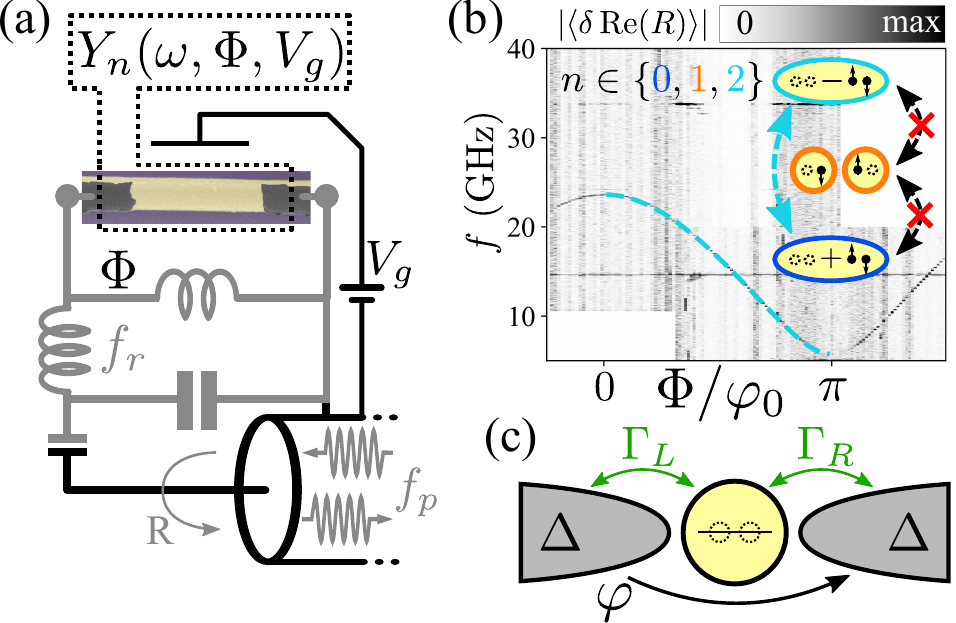}
\caption{ \label{fig:intro_schem}
(a) Schematic of the~\cqed~setup (description in main text), including a colorized scanning electron micrograph of a representative Josephson nanowire weak link (aluminum is dark gray, uncovered region is yellow and \SI{500}{\nano\meter} long in this image, but \SI{350}{\nano\meter} long in the device measured for this report). 
A probe tone $f_p$ reflects with a frequency-dependent reflection amplitude $R$ which is recorded to measure the system.
(b)  Two-tone spectroscopy as a function of drive frequency and phase bias revealed a single dispersing Andreev transition, with model fit (dashed cyan line, see Figure 2 for parameters) corresponding to the transition between the even parity states (inset schematic). 
Parity-switching transitions are forbidden (black arrows with red X). 
$n\in\{0,1,2\}$ label the eigenstates.
The horizontal lines near 15 and 34~\si{\giga\hertz} are higher harmonics of the resonator. 
(c) Minimal picture of a single-level quantum dot coupled to superconducting reservoirs with gap $\Delta$ and phase drop $\varphi$.
}
\end{figure}

The salient aspects of our \cqed~setup are depicted in Fig.~\ref{fig:intro_schem}(a). 
We grew an indium arsenide nanowire (micrograph) with two facets covered by epitaxial aluminum on which a $\SI{350}{\nano\meter}$ weak-link region was later uncovered.
The nanowire weak link is embedded in a superconducting loop that partially determines the inductance of the differential mode of a superconducting microwave resonator (gray).
An external coil biased with a dc current inserts magnetic flux $\Phi$ through the loop in order to phase bias the nanowire $\varphi \approx \Phi/\varphi_0$ ($\varphi_0=\hbar/2e$ is the reduced superconducting magnetic flux quantum).
The bare resonator frequency $\fbare =\SI{4.887}{\giga\hertz}$ was measured by depleting the weak link with a negative gate voltage $\Vgc$.

Discrete Andreev states were then introduced to the weak link by opening conduction channels via increasing $\Vgc$.
While doing this, we conducted two-tone spectroscopy measurements which detect microwave transitions between quantum states of the weak link.x
At low gate voltages, we detected gate "bias points" at which a single dispersing level was identifiable from the spectrum, such as in Fig.~\ref{fig:intro_schem}(b).
The discrete transition frequency depends strongly on phase, consistent with a transition between Andreev states at frequency $f_A$, characterized as the excitation of a pair of localized quasiparticles.
In similar microwave experiments, this transition is often compared with the short junction model~\cite{bretheau_localized_2013,janvier_coherent_2015,van_woerkom_microwave_2018,hays_direct_2018}, which requires $f_A = 2\Delta_\mathrm{Al}/h$ for $\Phi=0$.
However, the application of this model is inappropriate here since the measured transition frequency is well below twice the gap of the superconducting leads $f_A < \SI{25}{\giga\hertz} \ll 2\Delta_\mathrm{Al}/h \approx \SI{100}{\giga\hertz}$.

Resolution of this discrepancy requires a model that includes a nonzero dwell time for electrons in the weak link.
Here we focus on a minimal phenomenological model: a quantum dot with a single level coupled to two superconducting leads~\cite{beenakker_resonant_1992,wendin_josephson_1996,pillet_andreev_2010}, schematically shown in Fig.~\ref{fig:intro_schem}(c) \footnote{Models with weak link length of order the coherence length also exhibit dwell time. Such models admit additional doublets and add theoretical complications, see supplement V.E.}.
Within this model, dot states with zero and two electrons are hybridized due to the proximity effect between the dot and the leads, parameterized by the tunneling rates $\Gamma_{L,R}$. 
In the absence of the charging effect, these states would split symmetrically by $\pm E_A$ with respect to the one-electron state.
This splitting leads to the microwave transition frequency $h f_A = 2 E_A$.
As long as the proximity effect is weak, the Andreev states formed by this hybridization remain well-detached from the superconducting gap which leads to $f_{A}\ll 2\Delta_\mathrm{Al}/h$ consistent with the experiment.
States with a single electron at the dot are also necessarily present in the system.
However, when only a single level is present, we cannot probe them with two-tone spectroscopy which preserves fermion parity (inset of Fig.~\ref{fig:intro_schem}(b)).

\begin{figure*}
\includegraphics[width=2.0\columnwidth]{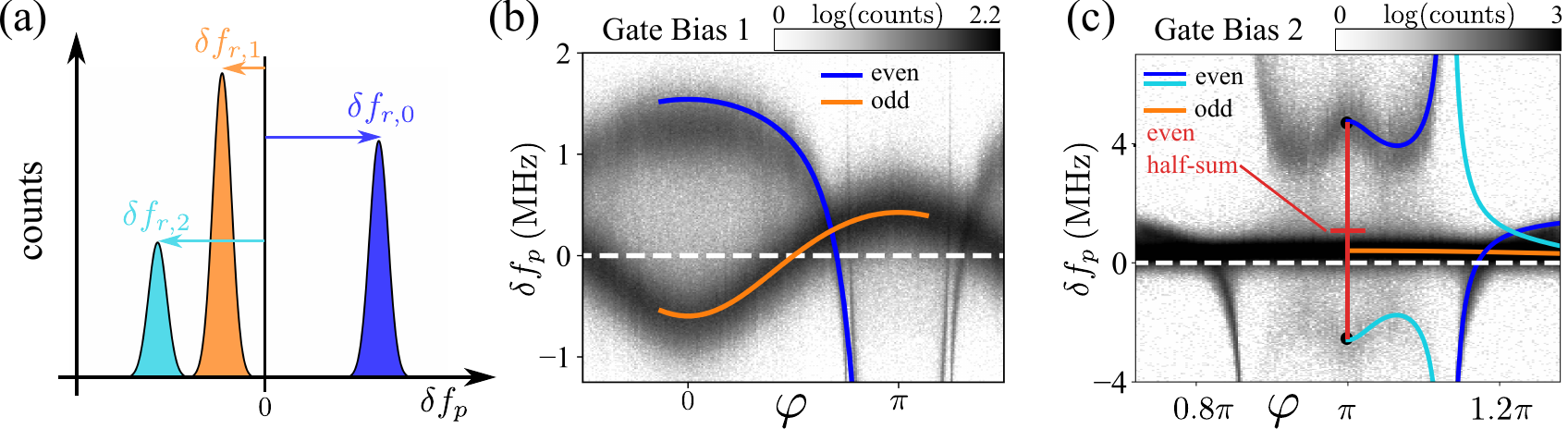}
\caption{ \label{fig:main_data}
(a) Schematic of the resonator frequency shift histogram: while sweeping probe frequency relative to the bare resonator frequency $\delta f_p = f_p - f_{r,b}$, we record counts when the reflection signal $R$ indicates a resonance. 
The magnitude of the frequency shift $\dfn$ is proportional to the admittance $Y_n$ of state $n$, and normalized counts under each peak give the probability of the state.
The widths of the distributions are the same, as they derive from the signal-to-noise ratio of the measurement (see supplement for details on the measurement).
(b) Resonator frequency shift histogram from the same bias point as Fig.~\ref{fig:intro_schem}(b). 
Solid lines are model fits of odd (orange) and even (blue) state dispersive shifts $\dfn$ relative to the bare resonator frequency $f_{r,b}$ (white dashed). 
The fit parameters in $\si{\giga\hertz}$ are $\D/h=28.6\pm0.3, \GL/h=6.1\pm0.03, \GR/h=10.4\pm0.06, U/h=21.6\pm0.8$, and a small offset of $\delta f_{\mathrm{off}} = (0.045\pm0.004) \si{\mega\hertz}$ to the resonator which is 2\% of the resonator linewidth of $\SI{2}{\mega\hertz}$.
See supplement for a discussion about the small value for $\D$, which we believe to arise from finite-length effects. 
(c)  Similar to (b) for bias point 2 in a narrow range of phase (see Fig. S8 for a larger phase range).
Black circles mark extracted $\dfn$ at $\varphi=\pi$, and the horizontal red bar marks the average of the even-parity frequencies $ \frac{1}{2} (\delta f_{r,0}+\delta f_{r,2})$ at $\varphi=\pi$, displaying clear violation of the half-sum rule $\frac{1}{2} (\delta f_{r,0}+\delta f_{r,2}) - \delta f_{r,1}=0$ (standard deviation is smaller than the red bar thickness).
The fit parameters in $\si{\giga\hertz}$ are $\D/h=25\pm6,\GL/h=7.0\pm1.6,\GR/h=8.1\pm1.8,U/h=74\pm9,\delta f_{\mathrm{off}}=(1.8\pm1.2)\times10^{-4}$.
}
\end{figure*}

We overcome this limitation with our \cqed~setup by measuring the discrete frequency shifts of the resonator.
The frequency shifts $\dfn$ are determined by the state-dependent admittance $Y_n$ of the weak link~\cite{park_adiabatic_2020,kurilovich_microwave_2021}: $\dfn \propto \mathrm{Im}(Y_n(2\pi \fbare))$ (see supplement for more on this). 
Here $n$ labels the many-body state of the link: $n=0,2$ are the two even parity states and $n=1$ is the spin-degenerate odd state, such that $n$ can be related to the number of quasiparticles in the weak link.

The admittance is given by $Y_n(\omega)= L_n^{-1}/(i \omega) + Y_{n,\mathrm{res}}(\omega)$. 
The first term describes the quasistatic response of the weak link and is determined by its inverse inductance, $L_n^{-1}=\varphi_0^{-2} \partial^2_\varphi E_n$. 
This contribution carries information about the unique energy-phase relations $E_n(\varphi)$ of each microscopic state, which is out of reach of two tone spectroscopy. 
The second term, $Y_{n, {\rm res}}(\omega)$, describes the resonant contribution to the admittance and is only appreciable close to the transition frequencies of the weak link in a given state.

Frequency shift histograms measured as a function of phase are shown for two gate bias points in Figure~\ref{fig:main_data}(b-c)~\footnote{Probe tone power corresponding to roughly 9 photons in the resonator.} (details on the overlaid fit curves are given further below). 
Fig.~\ref{fig:main_data}(b) corresponds to the same bias point as Fig.~\ref{fig:intro_schem}(b). 
The lowest-energy even state $n=0$ (blue) is identified by the strong dispersive shift of the resonator when the Andreev pair transition (Fig.~\ref{fig:main_data}(a)) approaches the resonator frequency near $\varphi=\pi$.
At the second bias point, Fig.~\ref{fig:main_data}(c), the Andreev transition crosses the resonator frequency leading to an anti-crossing-like feature for the even states. 
There, the higher-energy even state ($n=2$) is additionally visible (teal). 
At both bias points, an additional state is present in the data. 
The dispersive shift in this state does not have resonant signatures seen in the even states.
We thus identify it as the state with odd fermion parity ($n=1$). 

These measurements exhibit two qualitative discrepancies with the phenomenological dot model in its simplest limit~\cite{beenakker_resonant_1992,wendin_josephson_1996} of weak coupling to reservoirs and negligible charging energy.
First, the odd state $n=1$ exhibits a strong dispersion in the measurement.
In contrast, in the above limit the energies $E_n$ of the states $n\in\{0,1,2\}$ can be summarized by
\begin{equation}
    E_n = (n-1)E_A\label{oldthy}
\end{equation}
which has $E_1=0$.
Combined with the lack of transitions, this would result in $\delta f_{r,1}=0$. 
Note also that $\delta f_{r,1}$ is $\pi$-phase shifted relative to $\delta f_{r,0}$ which excludes a parallel channel interpretation.
The second qualitative feature is observed when all three possible occupations $n\in\{0,1,2\}$ are observed, as in Fig.~\ref{fig:main_data}(c).
There, the average frequency shift of the even states (the "half-sum" of $n=0,2$) is different from that of the odd state ($n=1$): $\frac{1}{2}(\delta f_{r,0}+\delta f_{r,2})-\delta f_{r,1} \approx \SI{0.6}{\mega\hertz} $~\footnote{This na{\"\i}ve comparison is in principle susceptible to corrections to the dispersive shifts because they are solutions of a transcendental equation which is inherently nonlinear in $Y(\omega)$ (see supplement). We estimate the corrections to be less than 10\% of the mismatch observed here. Furthermore, our probe power is small enough that frequency shifts due to nonlinearities are minimal.}.
This observation is indicated by the red lines and does not require comparison to a model.
Conventional noninteracting pictures for Andreev states (like the weak-coupling model described earlier) would predict zero such difference (i.e., the half-sum rule $\frac{1}{2}(\delta f_{r,0}+\delta f_{r,2})-\delta f_{r,1} = 0$) due to a particle-hole symmetry, as in Eq.~\eqref{oldthy}.  

How may we reconcile these two discrepancies?
First, the dispersion of the odd state may be accounted for by a phase-dispersing spectral continuum above the gap of the superconductor. 
The continuum is thus understood to produce a contribution $E_\mathrm{cont}$ common to all many-body states $n\in\{0,1,2\}$.
As we describe below, the presence of the appreciable continuum contribution to energy is a consequence of intermediate-strength tunnel coupling between the dot and the superconducting leads.
Second, the half-sum violation suggests that particle-hole symmetry is broken by a charging energy in the weak link.
These notions prompt us to qualitatively generalize the prior formulation to 
\begin{equation}
    E_n = E_\mathrm{cont} + (n-1)E_A + (n-1)^{2} U_A \label{newthy}
\end{equation}
where $U_A$ is a term related to charging energy that is phase-dependent since the Andreev eigenstates are not charge eigenstates.
Thus, both experimental discrepancies may be accounted for via inductive contributions to the dispersive shifts as follows from Eq.~\eqref{newthy}: $\partial^2_\varphi E_1=\partial^2_\varphi E_{\mathrm{cont}}$ and $\frac{1}{2}(\partial^2_\varphi E_0+ \partial^2_\varphi E_2)-\partial^2_\varphi E_1 = \partial^2_\varphi U_A$.

\begin{figure}
\includegraphics[width=1.0\columnwidth]{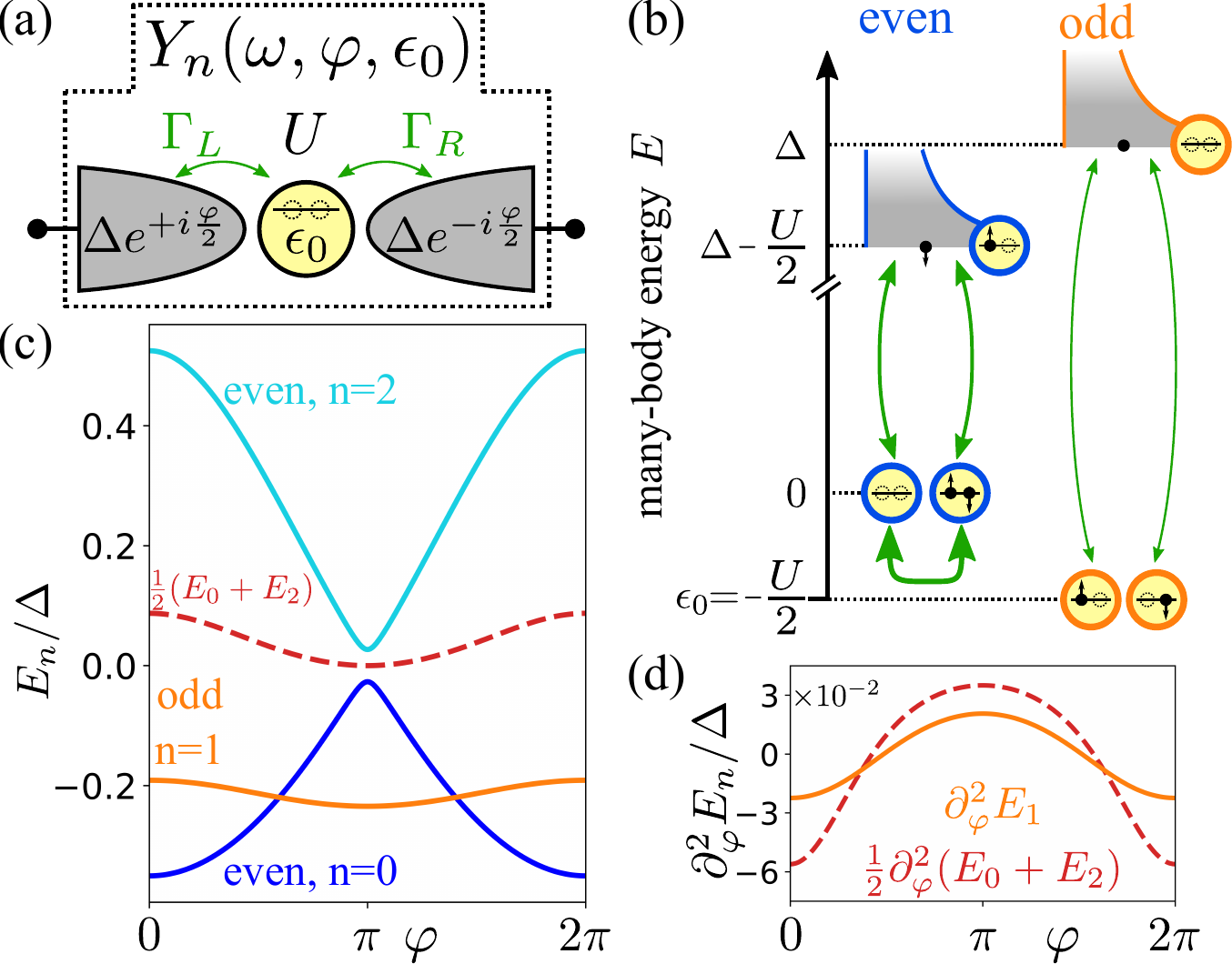}
\caption{ \label{fig:model}
(a) Diagram for the model of a single-level quantum dot weak link with state- and frequency-dependent admittance $Y_n(\omega)$. 
The dot has energy offset $\eo$, charging energy $U$, and coupling rates $\Gamma_{L,R}$ to the reservoirs on the left and right.
The reservoirs have the same pair potential $\Delta$ and a phase drop $\varphi$ between them.
(b) Energy of many-body states of the uncoupled level and reservoirs for $\eo=-U/2$. 
Double-headed arrows depict hybridization due to the couplings $\Gamma_{L,R}$, with line thickness an indicator of hybridization strength. 
The condition $\eo=-U/2$ allows the even manifold to exhibit resonant behavior.
(c) The energy dispersion $E_n(\varphi)$ of the four discrete many-body states (recall the odd state is spin degenerate) for $\GL/\D=0.32, \GR/\D=0.28, U=2(\GL+\GR), \eo=-U/2$. 
Coulomb interaction causes an energetic offset between the even and odd manifolds (red dashed vs orange).
(d) $\partial^2_\varphi E_n$ (proportional to inverse inductance) for the odd state is weaker than the half-sum of the even states -- a measurable distinction in the microwave response caused by charging. 
}
\end{figure}

Now, with the aid of Fig.~\ref{fig:model}, we build up a detailed model based on our recent work~\cite{kurilovich_microwave_2021} to which we will make a quantitative comparison with the data. 
We define a dot with a level of energy $\eo$ with respect to the chemical potential, on-site Coulomb interaction strength $U$, and coupling rates $\Gamma_{L,R}$ to the superconducting leads with pair potential $\Delta$ and phase difference $\varphi$, schematically shown in Fig.~\ref{fig:model}(a-b).
When isolated, the dot has four possible electron occupations that serve as basis states: zero and two electrons (even parity) as well as one electron (odd parity, spin degenerate). 
These basis states have energies $0$, $2\eo+U$, and $\eo$, respectively.
The reservoir couplings $\Gamma_{L,R}$ hybridize the even states, which develop a strong dispersion with phase $\varphi$, while the odd states are not coupled to each other.
The experimentally important configuration $\eo=-U/2$ is depicted in Fig.~\ref{fig:model}(b-d).
Here, the uncoupled even basis states are degenerate, and thus (provided $\GL\approx\GR$) the resulting Andreev states exhibit a weak anticrossing at phase $\varphi=\pi$, signifying resonant transmission~\cite{beenakker_resonant_1992,zazunov_andreev_2003}.

The couplings $\Gamma_{L,R}$ also hybridize the dot states with the reservoir continuum (vertical green arrows in Fig.~\ref{fig:model}(b)).
Consequently, the spectral continuum contributes $E_\mathrm{cont}(\varphi)$ to the energy-phase relations of all many-body states.
$E_\mathrm{cont}$ is $\pi$-phase shifted compared to the lowest even state.
Notably, the odd state dispersion (orange curve in Fig.~\ref{fig:model}(c)) is entirely due to the continuum within this model.

Finally, nonzero charging energy $U>0$ qualitatively influences the spectrum by raising the energy of the two-electron basis state, resulting in the term $(n-1)^2 U_A(\varphi)$ in Eq.~\eqref{newthy}.
This term may cause an odd ground state for some or all phases (Fig.~\ref{fig:model}(c)). 
Moreover, $U_A(\varphi)$ is $\pi$-shifted like $E_\mathrm{cont}$ (Fig.~\ref{fig:model}(d)).

We now return to the experimental data of Fig.~\ref{fig:main_data} to compare to this theoretical model.
The model includes Coulomb interaction perturbatively, assuming $U \ll \Delta + \Gamma$~\cite{kurilovich_microwave_2021}.
At each gate bias point, we performed a least-squares fit to all available dispersive data ($\dfn$ and the transition $f_A$) simultaneously. 
The bias points are gate voltage $\Vgc$ ``sweet spots'' where the transition frequency was at a minimum, thus minimizing charge noise (see supplement). 
In our model, such a sweet spot corresponds to the condition $\eo\equiv-U/2$ for a gate-controlled energy $\eo$.

The fitted model is overlaid on the data (Fig.~\ref{fig:intro_schem}(b) and Fig.~\ref{fig:main_data}(b) for gate bias point 1, Fig.~\ref{fig:main_data}(c) for gate bias point 2).
We begin with bias point 1. 
The model quantitatively reproduces the $\pi$-phase shifted response of the odd parity state (orange) resulting from the continuum.
The continuum contribution to the even state $n=0$ is also important: it results in a lower admittance (smaller frequency shift) than would otherwise be expected, an important effect when incorporating such weak links into microwave circuits~\cite{bargerbos_observation_2020,kringhoj_suppressed_2020,metzger_circuit-qed_2021}.

The model also accounts for the violation of the half-sum rule via the charging energy, as shown in Fig.~\ref{fig:main_data}(c). 
The theory correctly accounts for the sign of the effect $\frac{1}{2}(\delta f_{r,0}+\delta f_{r,2}) > \delta f_{r,1}$.
However, the fit in Fig.~\ref{fig:main_data}(c) gives $U \gtrsim \D+\Gamma$, which is beyond the validity of our perturbative theory (see Fig.~\ref{fig:main_data} caption and~\cite{kurilovich_microwave_2021}).
Nonetheless, our simple model allows us to see that charging effects in the weak link are appreciable~\footnote{We note that a naive estimation based on the dimensions of just the uncovered weak link region gives of order $U \sim \SI{200}{\giga\hertz}$, whereas our fits produce values of $U$ that are up to an order of magnitude smaller. This suggests strong screening by the epitaxial aluminum leads.}.

\begin{figure}
\includegraphics[width=0.99\columnwidth]{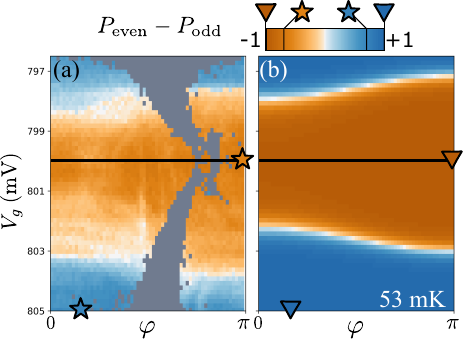}
\caption{ \label{fig:stability}
The difference in the probability of occupation of the even $n=0$ and odd $n=1$ states $P_{\mathrm{even}}-P_{\mathrm{odd}}$ (i.e. the fermion parity polarization) as a function of $\varphi$ and $\Vgc$ near gate bias point 2. 
Panel (a) is the experiment and (b) is a thermal equilibrium prediction within our model for the energies. 
Both panels have the same pixel size and use the same color scale.
We assume $\eo\propto \Vgc$ for simplicity (see supplement for calibration).
(a) Gray regions denote where $\dfn$ were too similar to obtain reliable information. 
Orange and blue stars indicate the extremal experimental polarizations that are also marked on the colorbar. 
(b) We use $T=\SI{53}{\milli\kelvin}$ for the thermal equilibrium prediction (see supplement for calibration).
All other parameters were taken from the fit in Fig.~\ref{fig:main_data}(c), which was taken along the black line. 
}
\end{figure}

As mentioned, charging effects of this strength can overcome superconducting pairing to produce an odd parity ground state. 
Within the model, the odd state is energetically favored for a range of phase and gate voltage centered at $\varphi=\pi$ and $\eo=-U/2$. 
In order to check this in the experiment, we quantified the difference in the probability of the fermion parities $P_\mathrm{even}-P_\mathrm{odd}$ (the polarization) as a function of phase $\varphi$ and gate voltage $\Vgc$ for bias point 2, shown in Fig.~\ref{fig:stability}(a) (see supplement for details).
The sweet spot at $\Vgc=\SI{800}{\milli\volt}$ (same as Fig.~\ref{fig:main_data}(c)) exhibits majority odd population (orange) for all phases.
Detuning the gate voltage $\Vgc$ from this point results in a transition (white) from majority-odd (orange) to majority-even (blue).
The transition voltage increases when tuning $\varphi$ from $0$ to $\pi$, indeed suggestive of a more stable odd state at $\varphi=\pi$.
Together, these results are consistent with a 0-$\pi$ transition for the weak link, as would be expected from the energies of the interacting quantum dot model. 
To the best of our knowledge, this is the first time that such a phase diagram has been measured in a microwave experiment.

We can further compare with the predictions of thermal equilibrium using the state energies predicted by our model, shown Fig.~\ref{fig:stability}(b).
For this, we used an independently calibrated weak link temperature of $\SI{53}{\milli\kelvin}$ based on the relative probability of the same-parity states $n=0,2$ observed near phase $\varphi=\pi$ (see supplement). 
We find a similar pattern for the sign of the polarization as the experiment, but a striking difference is evident in the magnitude of polarization, indicated by the darkness of the colors. 
Most of the theoretical phase diagram exhibits full polarization $1 - |P_\mathrm{even}-P_\mathrm{odd}| \ll 1 $ because the typical fermion parity energy difference is much larger than the inferred temperature $|E_0-E_1|/k_\mathrm{B} \gg \SI{53}{\milli\kelvin}$ ($k_\mathrm{B}$ is Boltzmann's constant).
The experimental weak link, however, never exceeds a polarization of 0.7, which, taken at face value, would require temperature $>\SI{500}{\milli\kelvin}$, over an order of magnitude larger than that seen in the even-states. 
Quasiparticles in superconducting mesoscopic devices are known to be out of equilibrium and so parity dynamics indeed may not follow from the expectations of thermodynamic equilibrium~\cite{glazman_bogoliubov_2021}.
Our data points to the need for a model of nonequilibrium fermion parity dynamics to quantitatively explain the variation of the polarization with phase bias. 
We leave this task for future work.

In closing, we employed a \cqed~setup to isolate and quantify the microwave response of the quantum many-body states in a superconducting weak link. 
We identified and elucidated two important features in the dispersion of the microwave response functions. 
One is a violation of a particle-hole symmetry that results from the influence of charging effects. 
The second is a substantial phase dispersion common to all states which is consistent with the presence of a phase-dispersing spectral continuum.
In addition, we observed switches of the fermion parity polarization when varying gate voltage and phase bias that are consistent with a 0-$\pi$ transition.  
Mysteries to be resolved in the future include the incompleteness of the fermion parity polarization, the consistent overestimation of the $n=0$ admittance near $\varphi=0$, as well as the role of yet-unidentified high-frequency transitions in spectroscopy at gate bias 2 (see Fig.~(S8)).
Overall, our results demonstrate that the physics of charging energy and the spectral continuum cannot be ignored in the description of weak links hosting Andreev levels in superconductor-semiconductor Josephson devices and open a path to their detailed study.

\begin{acknowledgments}
\emph{Acknowledgements:}
We would like to acknowledge two other experimental teams pursuing related
research on Coulomb interaction in Andreev states. 
We thank F. J. Matute Canadas, C. Metzger, S. Park, L. Tosi, M. F. Goffman, C. Urbina, H. Pothier, A. Levy-Yeyati for sharing their manuscript~\cite{matute_canadas_signatures_2021}, for useful comments on ours, and for fruitful discussions.
We thank A. Bargerbos, M. Pita-Vidal,  R. Zitko, R. Aguado, A. Kou, and B. van Heck for sharing their data~\cite{bargerbos_singlet-doublet_nodate} and for helpful comments and discussions.
We also thank M. Houzet, J. Vayrynen, and L. Bretheau for helpful discussions and L. Frunzio for assistance with device fabrication.
This work is supported by the ARO under Grant Number W911NF-18-1-0212. 
V.D.K., L.I.G. acknowledge the support of NSF DMR-2002275.
D.B. acknowledges support by Netherlands Organisation for Scientific Research (NWO) and Microsoft Corporation Station Q. 
P.K. acknowledges support from Microsoft Quantum and the European Research Council under the grant agreement No.716655 (HEMs-DAM).
J.N. acknowledges support from the Danish National Research Foundation.
Some of the authors acknowledge the European Union’s Horizon 2020 research and innovation programme for financial support: A.G received funding from the European Research Council, grant no. 804988 (SiMS), and A.G. and J.N. further acknowledge grant no. 828948 (AndQC) and QuantERA project no. 127900 (SuperTOP). 
We thank Will Oliver and Lincoln Laboratories for providing the traveling wave parametric amplifier used in this experiment.
Facilities use was supported by YINQE and the Yale SEAS cleanroom.
We also acknowledge the Yale Quantum Institute.
\end{acknowledgments}

\paragraph*{Author contributions}
P.K. and J.N. developed the nanowire materials.
V.F., M.H., N.F., D.B., S.D., and A.G. designed the experimental setup.
V.F., D.B., and S.D. fabricated the nanowire and resonator device.
V.F. performed the measurements with feedback from P.D.K., M.H., V.D.K., N.F., and S.D.
V.F., P.D.K., V.D.K., M.H., T.C., N.F., S.D., L.I.G., and M.H.D. analyzed the data.
V.F., P.D.K., V.D.K., L.I.G., and M.H.D. wrote the manuscript with feedback from all authors.

\bibliography{references}

\begin{thebibliography}{45}%
\makeatletter
\providecommand \@ifxundefined [1]{%
 \@ifx{#1\undefined}
}%
\providecommand \@ifnum [1]{%
 \ifnum #1\expandafter \@firstoftwo
 \else \expandafter \@secondoftwo
 \fi
}%
\providecommand \@ifx [1]{%
 \ifx #1\expandafter \@firstoftwo
 \else \expandafter \@secondoftwo
 \fi
}%
\providecommand \natexlab [1]{#1}%
\providecommand \enquote  [1]{``#1''}%
\providecommand \bibnamefont  [1]{#1}%
\providecommand \bibfnamefont [1]{#1}%
\providecommand \citenamefont [1]{#1}%
\providecommand \href@noop [0]{\@secondoftwo}%
\providecommand \href [0]{\begingroup \@sanitize@url \@href}%
\providecommand \@href[1]{\@@startlink{#1}\@@href}%
\providecommand \@@href[1]{\endgroup#1\@@endlink}%
\providecommand \@sanitize@url [0]{\catcode `\\12\catcode `\$12\catcode
  `\&12\catcode `\#12\catcode `\^12\catcode `\_12\catcode `\%12\relax}%
\providecommand \@@startlink[1]{}%
\providecommand \@@endlink[0]{}%
\providecommand \url  [0]{\begingroup\@sanitize@url \@url }%
\providecommand \@url [1]{\endgroup\@href {#1}{\urlprefix }}%
\providecommand \urlprefix  [0]{URL }%
\providecommand \Eprint [0]{\href }%
\providecommand \doibase [0]{https://doi.org/}%
\providecommand \selectlanguage [0]{\@gobble}%
\providecommand \bibinfo  [0]{\@secondoftwo}%
\providecommand \bibfield  [0]{\@secondoftwo}%
\providecommand \translation [1]{[#1]}%
\providecommand \BibitemOpen [0]{}%
\providecommand \bibitemStop [0]{}%
\providecommand \bibitemNoStop [0]{.\EOS\space}%
\providecommand \EOS [0]{\spacefactor3000\relax}%
\providecommand \BibitemShut  [1]{\csname bibitem#1\endcsname}%
\let\auto@bib@innerbib\@empty
\bibitem [{\citenamefont {Bretheau}(2013)}]{bretheau_localized_2013}%
  \BibitemOpen
  \bibfield  {author} {\bibinfo {author} {\bibfnamefont {L.}~\bibnamefont
  {Bretheau}},\ }\emph {\bibinfo {title} {Localized {Excitations} in
  {Superconducting} {Atomic} {Contacts}: {Probing} the {Andreev} {Doublet}}},\
  \href {https://pastel.archives-ouvertes.fr/pastel-00862029/document}
  {\bibinfo {type} {Thesis}},\ \bibinfo  {school} {Ecole Polytechnique}
  (\bibinfo {year} {2013})\BibitemShut {NoStop}%
\bibitem [{\citenamefont {Janvier}\ \emph {et~al.}(2015)\citenamefont
  {Janvier}, \citenamefont {Tosi}, \citenamefont {Bretheau}, \citenamefont
  {Girit}, \citenamefont {Stern}, \citenamefont {Bertet}, \citenamefont
  {Joyez}, \citenamefont {Vion}, \citenamefont {Esteve}, \citenamefont
  {Goffman}, \citenamefont {Pothier},\ and\ \citenamefont
  {Urbina}}]{janvier_coherent_2015}%
  \BibitemOpen
  \bibfield  {author} {\bibinfo {author} {\bibfnamefont {C.}~\bibnamefont
  {Janvier}}, \bibinfo {author} {\bibfnamefont {L.}~\bibnamefont {Tosi}},
  \bibinfo {author} {\bibfnamefont {L.}~\bibnamefont {Bretheau}}, \bibinfo
  {author} {\bibfnamefont {C.~O.}\ \bibnamefont {Girit}}, \bibinfo {author}
  {\bibfnamefont {M.}~\bibnamefont {Stern}}, \bibinfo {author} {\bibfnamefont
  {P.}~\bibnamefont {Bertet}}, \bibinfo {author} {\bibfnamefont
  {P.}~\bibnamefont {Joyez}}, \bibinfo {author} {\bibfnamefont
  {D.}~\bibnamefont {Vion}}, \bibinfo {author} {\bibfnamefont {D.}~\bibnamefont
  {Esteve}}, \bibinfo {author} {\bibfnamefont {M.~F.}\ \bibnamefont {Goffman}},
  \bibinfo {author} {\bibfnamefont {H.}~\bibnamefont {Pothier}},\ and\ \bibinfo
  {author} {\bibfnamefont {C.}~\bibnamefont {Urbina}},\ }\bibfield  {title}
  {\bibinfo {title} {Coherent manipulation of {Andreev} states in
  superconducting atomic contacts},\ }\href
  {https://doi.org/10.1126/science.aab2179} {\bibfield  {journal} {\bibinfo
  {journal} {Science}\ }\textbf {\bibinfo {volume} {349}},\ \bibinfo {pages}
  {1199} (\bibinfo {year} {2015})}\BibitemShut {NoStop}%
\bibitem [{\citenamefont {van Woerkom}\ \emph {et~al.}(2017)\citenamefont {van
  Woerkom}, \citenamefont {Proutski}, \citenamefont {van Heck}, \citenamefont
  {Bouman}, \citenamefont {Väyrynen}, \citenamefont {Glazman}, \citenamefont
  {Krogstrup}, \citenamefont {Nygård}, \citenamefont {Kouwenhoven},\ and\
  \citenamefont {Geresdi}}]{van_woerkom_microwave_2017}%
  \BibitemOpen
  \bibfield  {author} {\bibinfo {author} {\bibfnamefont {D.~J.}\ \bibnamefont
  {van Woerkom}}, \bibinfo {author} {\bibfnamefont {A.}~\bibnamefont
  {Proutski}}, \bibinfo {author} {\bibfnamefont {B.}~\bibnamefont {van Heck}},
  \bibinfo {author} {\bibfnamefont {D.}~\bibnamefont {Bouman}}, \bibinfo
  {author} {\bibfnamefont {J.~I.}\ \bibnamefont {Väyrynen}}, \bibinfo {author}
  {\bibfnamefont {L.~I.}\ \bibnamefont {Glazman}}, \bibinfo {author}
  {\bibfnamefont {P.}~\bibnamefont {Krogstrup}}, \bibinfo {author}
  {\bibfnamefont {J.}~\bibnamefont {Nygård}}, \bibinfo {author} {\bibfnamefont
  {L.~P.}\ \bibnamefont {Kouwenhoven}},\ and\ \bibinfo {author} {\bibfnamefont
  {A.}~\bibnamefont {Geresdi}},\ }\bibfield  {title} {\bibinfo {title}
  {Microwave spectroscopy of spinful {Andreev} bound states in ballistic
  semiconductor {Josephson} junctions},\ }\href
  {https://doi.org/10.1038/nphys4150} {\bibfield  {journal} {\bibinfo
  {journal} {Nature Physics}\ }\textbf {\bibinfo {volume} {13}},\ \bibinfo
  {pages} {876} (\bibinfo {year} {2017})}\BibitemShut {NoStop}%
\bibitem [{\citenamefont {Hays}\ \emph {et~al.}(2018)\citenamefont {Hays},
  \citenamefont {de~Lange}, \citenamefont {Serniak}, \citenamefont {van
  Woerkom}, \citenamefont {Bouman}, \citenamefont {Krogstrup}, \citenamefont
  {Nygård}, \citenamefont {Geresdi},\ and\ \citenamefont
  {Devoret}}]{hays_direct_2018}%
  \BibitemOpen
  \bibfield  {author} {\bibinfo {author} {\bibfnamefont {M.}~\bibnamefont
  {Hays}}, \bibinfo {author} {\bibfnamefont {G.}~\bibnamefont {de~Lange}},
  \bibinfo {author} {\bibfnamefont {K.}~\bibnamefont {Serniak}}, \bibinfo
  {author} {\bibfnamefont {D.}~\bibnamefont {van Woerkom}}, \bibinfo {author}
  {\bibfnamefont {D.}~\bibnamefont {Bouman}}, \bibinfo {author} {\bibfnamefont
  {P.}~\bibnamefont {Krogstrup}}, \bibinfo {author} {\bibfnamefont
  {J.}~\bibnamefont {Nygård}}, \bibinfo {author} {\bibfnamefont
  {A.}~\bibnamefont {Geresdi}},\ and\ \bibinfo {author} {\bibfnamefont
  {M.}~\bibnamefont {Devoret}},\ }\bibfield  {title} {\bibinfo {title} {Direct
  {Microwave} {Measurement} of {Andreev}-{Bound}-{State} {Dynamics} in a
  {Semiconductor}-{Nanowire} {Josephson} {Junction}},\ }\href
  {https://doi.org/10.1103/PhysRevLett.121.047001} {\bibfield  {journal}
  {\bibinfo  {journal} {Physical Review Letters}\ }\textbf {\bibinfo {volume}
  {121}},\ \bibinfo {pages} {047001} (\bibinfo {year} {2018})}\BibitemShut
  {NoStop}%
\bibitem [{\citenamefont {Tosi}\ \emph {et~al.}(2019)\citenamefont {Tosi},
  \citenamefont {Metzger}, \citenamefont {Goffman}, \citenamefont {Urbina},
  \citenamefont {Pothier}, \citenamefont {Park}, \citenamefont {Yeyati},
  \citenamefont {Nygård},\ and\ \citenamefont
  {Krogstrup}}]{tosi_spin-orbit_2019}%
  \BibitemOpen
  \bibfield  {author} {\bibinfo {author} {\bibfnamefont {L.}~\bibnamefont
  {Tosi}}, \bibinfo {author} {\bibfnamefont {C.}~\bibnamefont {Metzger}},
  \bibinfo {author} {\bibfnamefont {M.}~\bibnamefont {Goffman}}, \bibinfo
  {author} {\bibfnamefont {C.}~\bibnamefont {Urbina}}, \bibinfo {author}
  {\bibfnamefont {H.}~\bibnamefont {Pothier}}, \bibinfo {author} {\bibfnamefont
  {S.}~\bibnamefont {Park}}, \bibinfo {author} {\bibfnamefont {A.~L.}\
  \bibnamefont {Yeyati}}, \bibinfo {author} {\bibfnamefont {J.}~\bibnamefont
  {Nygård}},\ and\ \bibinfo {author} {\bibfnamefont {P.}~\bibnamefont
  {Krogstrup}},\ }\bibfield  {title} {\bibinfo {title} {Spin-{Orbit}
  {Splitting} of {Andreev} {States} {Revealed} by {Microwave} {Spectroscopy}},\
  }\href {https://doi.org/10.1103/PhysRevX.9.011010} {\bibfield  {journal}
  {\bibinfo  {journal} {Physical Review X}\ }\textbf {\bibinfo {volume} {9}},\
  \bibinfo {pages} {011010} (\bibinfo {year} {2019})}\BibitemShut {NoStop}%
\bibitem [{\citenamefont {Hays}\ \emph {et~al.}(2020)\citenamefont {Hays},
  \citenamefont {Fatemi}, \citenamefont {Serniak}, \citenamefont {Bouman},
  \citenamefont {Diamond}, \citenamefont {de~Lange}, \citenamefont {Krogstrup},
  \citenamefont {Nygård}, \citenamefont {Geresdi},\ and\ \citenamefont
  {Devoret}}]{hays_continuous_2020}%
  \BibitemOpen
  \bibfield  {author} {\bibinfo {author} {\bibfnamefont {M.}~\bibnamefont
  {Hays}}, \bibinfo {author} {\bibfnamefont {V.}~\bibnamefont {Fatemi}},
  \bibinfo {author} {\bibfnamefont {K.}~\bibnamefont {Serniak}}, \bibinfo
  {author} {\bibfnamefont {D.}~\bibnamefont {Bouman}}, \bibinfo {author}
  {\bibfnamefont {S.}~\bibnamefont {Diamond}}, \bibinfo {author} {\bibfnamefont
  {G.}~\bibnamefont {de~Lange}}, \bibinfo {author} {\bibfnamefont
  {P.}~\bibnamefont {Krogstrup}}, \bibinfo {author} {\bibfnamefont
  {J.}~\bibnamefont {Nygård}}, \bibinfo {author} {\bibfnamefont
  {A.}~\bibnamefont {Geresdi}},\ and\ \bibinfo {author} {\bibfnamefont {M.~H.}\
  \bibnamefont {Devoret}},\ }\bibfield  {title} {\bibinfo {title} {Continuous
  monitoring of a trapped superconducting spin},\ }\href
  {https://doi.org/10.1038/s41567-020-0952-3} {\bibfield  {journal} {\bibinfo
  {journal} {Nature Physics}\ }\textbf {\bibinfo {volume} {16}},\ \bibinfo
  {pages} {1103} (\bibinfo {year} {2020})}\BibitemShut {NoStop}%
\bibitem [{\citenamefont {Hays}\ \emph {et~al.}(2021)\citenamefont {Hays},
  \citenamefont {Fatemi}, \citenamefont {Bouman}, \citenamefont {Cerrillo},
  \citenamefont {Diamond}, \citenamefont {Serniak}, \citenamefont {Connolly},
  \citenamefont {Krogstrup}, \citenamefont {Nygård}, \citenamefont {Yeyati},
  \citenamefont {Geresdi},\ and\ \citenamefont {Devoret}}]{hays_coherent_2021}%
  \BibitemOpen
  \bibfield  {author} {\bibinfo {author} {\bibfnamefont {M.}~\bibnamefont
  {Hays}}, \bibinfo {author} {\bibfnamefont {V.}~\bibnamefont {Fatemi}},
  \bibinfo {author} {\bibfnamefont {D.}~\bibnamefont {Bouman}}, \bibinfo
  {author} {\bibfnamefont {J.}~\bibnamefont {Cerrillo}}, \bibinfo {author}
  {\bibfnamefont {S.}~\bibnamefont {Diamond}}, \bibinfo {author} {\bibfnamefont
  {K.}~\bibnamefont {Serniak}}, \bibinfo {author} {\bibfnamefont
  {T.}~\bibnamefont {Connolly}}, \bibinfo {author} {\bibfnamefont
  {P.}~\bibnamefont {Krogstrup}}, \bibinfo {author} {\bibfnamefont
  {J.}~\bibnamefont {Nygård}}, \bibinfo {author} {\bibfnamefont {A.~L.}\
  \bibnamefont {Yeyati}}, \bibinfo {author} {\bibfnamefont {A.}~\bibnamefont
  {Geresdi}},\ and\ \bibinfo {author} {\bibfnamefont {M.~H.}\ \bibnamefont
  {Devoret}},\ }\bibfield  {title} {\bibinfo {title} {Coherent manipulation of
  an {Andreev} spin qubit},\ }\href {https://doi.org/10.1126/science.abf0345}
  {\bibfield  {journal} {\bibinfo  {journal} {Science}\ }\textbf {\bibinfo
  {volume} {373}},\ \bibinfo {pages} {430} (\bibinfo {year}
  {2021})}\BibitemShut {NoStop}%
\bibitem [{\citenamefont {Kos}\ \emph {et~al.}(2013)\citenamefont {Kos},
  \citenamefont {Nigg},\ and\ \citenamefont
  {Glazman}}]{kos_frequency-dependent_2013}%
  \BibitemOpen
  \bibfield  {author} {\bibinfo {author} {\bibfnamefont {F.}~\bibnamefont
  {Kos}}, \bibinfo {author} {\bibfnamefont {S.~E.}\ \bibnamefont {Nigg}},\ and\
  \bibinfo {author} {\bibfnamefont {L.~I.}\ \bibnamefont {Glazman}},\
  }\bibfield  {title} {\bibinfo {title} {Frequency-dependent admittance of a
  short superconducting weak link},\ }\href
  {https://doi.org/10.1103/PhysRevB.87.174521} {\bibfield  {journal} {\bibinfo
  {journal} {Physical Review B}\ }\textbf {\bibinfo {volume} {87}},\ \bibinfo
  {pages} {174521} (\bibinfo {year} {2013})}\BibitemShut {NoStop}%
\bibitem [{\citenamefont {Park}\ and\ \citenamefont
  {Yeyati}(2017)}]{park_andreev_2017}%
  \BibitemOpen
  \bibfield  {author} {\bibinfo {author} {\bibfnamefont {S.}~\bibnamefont
  {Park}}\ and\ \bibinfo {author} {\bibfnamefont {A.~L.}\ \bibnamefont
  {Yeyati}},\ }\bibfield  {title} {\bibinfo {title} {Andreev spin qubits in
  multichannel {Rashba} nanowires},\ }\href
  {https://doi.org/10.1103/PhysRevB.96.125416} {\bibfield  {journal} {\bibinfo
  {journal} {Physical Review B}\ }\textbf {\bibinfo {volume} {96}},\ \bibinfo
  {pages} {125416} (\bibinfo {year} {2017})}\BibitemShut {NoStop}%
\bibitem [{\citenamefont {Park}\ \emph {et~al.}(2020)\citenamefont {Park},
  \citenamefont {Metzger}, \citenamefont {Tosi}, \citenamefont {Goffman},
  \citenamefont {Urbina}, \citenamefont {Pothier},\ and\ \citenamefont
  {Yeyati}}]{park_adiabatic_2020}%
  \BibitemOpen
  \bibfield  {author} {\bibinfo {author} {\bibfnamefont {S.}~\bibnamefont
  {Park}}, \bibinfo {author} {\bibfnamefont {C.}~\bibnamefont {Metzger}},
  \bibinfo {author} {\bibfnamefont {L.}~\bibnamefont {Tosi}}, \bibinfo {author}
  {\bibfnamefont {M.}~\bibnamefont {Goffman}}, \bibinfo {author} {\bibfnamefont
  {C.}~\bibnamefont {Urbina}}, \bibinfo {author} {\bibfnamefont
  {H.}~\bibnamefont {Pothier}},\ and\ \bibinfo {author} {\bibfnamefont {A.~L.}\
  \bibnamefont {Yeyati}},\ }\bibfield  {title} {\bibinfo {title} {From
  {Adiabatic} to {Dispersive} {Readout} of {Quantum} {Circuits}},\ }\href
  {https://doi.org/10.1103/PhysRevLett.125.077701} {\bibfield  {journal}
  {\bibinfo  {journal} {Physical Review Letters}\ }\textbf {\bibinfo {volume}
  {125}},\ \bibinfo {pages} {077701} (\bibinfo {year} {2020})}\BibitemShut
  {NoStop}%
\bibitem [{\citenamefont {Buitelaar}\ \emph {et~al.}(2003)\citenamefont
  {Buitelaar}, \citenamefont {Belzig}, \citenamefont {Nussbaumer},
  \citenamefont {Babić}, \citenamefont {Bruder},\ and\ \citenamefont
  {Schönenberger}}]{buitelaar_multiple_2003}%
  \BibitemOpen
  \bibfield  {author} {\bibinfo {author} {\bibfnamefont {M.~R.}\ \bibnamefont
  {Buitelaar}}, \bibinfo {author} {\bibfnamefont {W.}~\bibnamefont {Belzig}},
  \bibinfo {author} {\bibfnamefont {T.}~\bibnamefont {Nussbaumer}}, \bibinfo
  {author} {\bibfnamefont {B.}~\bibnamefont {Babić}}, \bibinfo {author}
  {\bibfnamefont {C.}~\bibnamefont {Bruder}},\ and\ \bibinfo {author}
  {\bibfnamefont {C.}~\bibnamefont {Schönenberger}},\ }\bibfield  {title}
  {\bibinfo {title} {Multiple {Andreev} {Reflections} in a {Carbon} {Nanotube}
  {Quantum} {Dot}},\ }\href {https://doi.org/10.1103/PhysRevLett.91.057005}
  {\bibfield  {journal} {\bibinfo  {journal} {Physical Review Letters}\
  }\textbf {\bibinfo {volume} {91}},\ \bibinfo {pages} {057005} (\bibinfo
  {year} {2003})}\BibitemShut {NoStop}%
\bibitem [{\citenamefont {van Dam}\ \emph {et~al.}(2006)\citenamefont {van
  Dam}, \citenamefont {Nazarov}, \citenamefont {Bakkers}, \citenamefont
  {De~Franceschi},\ and\ \citenamefont
  {Kouwenhoven}}]{van_dam_supercurrent_2006}%
  \BibitemOpen
  \bibfield  {author} {\bibinfo {author} {\bibfnamefont {J.~A.}\ \bibnamefont
  {van Dam}}, \bibinfo {author} {\bibfnamefont {Y.~V.}\ \bibnamefont
  {Nazarov}}, \bibinfo {author} {\bibfnamefont {E.~P. A.~M.}\ \bibnamefont
  {Bakkers}}, \bibinfo {author} {\bibfnamefont {S.}~\bibnamefont
  {De~Franceschi}},\ and\ \bibinfo {author} {\bibfnamefont {L.~P.}\
  \bibnamefont {Kouwenhoven}},\ }\bibfield  {title} {\bibinfo {title}
  {Supercurrent reversal in quantum dots},\ }\href
  {https://doi.org/10.1038/nature05018} {\bibfield  {journal} {\bibinfo
  {journal} {Nature}\ }\textbf {\bibinfo {volume} {442}},\ \bibinfo {pages}
  {667} (\bibinfo {year} {2006})}\BibitemShut {NoStop}%
\bibitem [{\citenamefont {Jorgensen}\ \emph {et~al.}(2007)\citenamefont
  {Jorgensen}, \citenamefont {Novotny}, \citenamefont {Grove-Rasmussen},
  \citenamefont {Flensberg},\ and\ \citenamefont
  {Lindelof}}]{jorgensen_critical_2007}%
  \BibitemOpen
  \bibfield  {author} {\bibinfo {author} {\bibfnamefont {H.~I.}\ \bibnamefont
  {Jorgensen}}, \bibinfo {author} {\bibfnamefont {T.}~\bibnamefont {Novotny}},
  \bibinfo {author} {\bibfnamefont {K.}~\bibnamefont {Grove-Rasmussen}},
  \bibinfo {author} {\bibfnamefont {K.}~\bibnamefont {Flensberg}},\ and\
  \bibinfo {author} {\bibfnamefont {P.~E.}\ \bibnamefont {Lindelof}},\
  }\bibfield  {title} {\bibinfo {title} {Critical {Current} 0-pi {Transition}
  in {Designed} {Josephson} {Quantum} {Dot} {Junctions}},\ }\href
  {https://doi.org/10.1021/nl071152w} {\bibfield  {journal} {\bibinfo
  {journal} {Nano Letters}\ }\textbf {\bibinfo {volume} {7}},\ \bibinfo {pages}
  {2441} (\bibinfo {year} {2007})}\BibitemShut {NoStop}%
\bibitem [{\citenamefont {Pillet}\ \emph {et~al.}(2010)\citenamefont {Pillet},
  \citenamefont {Quay}, \citenamefont {Morfin}, \citenamefont {Bena},
  \citenamefont {Yeyati},\ and\ \citenamefont {Joyez}}]{pillet_andreev_2010}%
  \BibitemOpen
  \bibfield  {author} {\bibinfo {author} {\bibfnamefont {J.-D.}\ \bibnamefont
  {Pillet}}, \bibinfo {author} {\bibfnamefont {C.~H.~L.}\ \bibnamefont {Quay}},
  \bibinfo {author} {\bibfnamefont {P.}~\bibnamefont {Morfin}}, \bibinfo
  {author} {\bibfnamefont {C.}~\bibnamefont {Bena}}, \bibinfo {author}
  {\bibfnamefont {A.~L.}\ \bibnamefont {Yeyati}},\ and\ \bibinfo {author}
  {\bibfnamefont {P.}~\bibnamefont {Joyez}},\ }\bibfield  {title} {\bibinfo
  {title} {Andreev bound states in supercurrent-carrying carbon nanotubes
  revealed},\ }\href {https://doi.org/10.1038/nphys1811} {\bibfield  {journal}
  {\bibinfo  {journal} {Nature Physics}\ }\textbf {\bibinfo {volume} {6}},\
  \bibinfo {pages} {965} (\bibinfo {year} {2010})}\BibitemShut {NoStop}%
\bibitem [{\citenamefont {Li}\ \emph {et~al.}(2017)\citenamefont {Li},
  \citenamefont {Kang}, \citenamefont {Caroff},\ and\ \citenamefont
  {Xu}}]{li_0-pi_2017}%
  \BibitemOpen
  \bibfield  {author} {\bibinfo {author} {\bibfnamefont {S.}~\bibnamefont
  {Li}}, \bibinfo {author} {\bibfnamefont {N.}~\bibnamefont {Kang}}, \bibinfo
  {author} {\bibfnamefont {P.}~\bibnamefont {Caroff}},\ and\ \bibinfo {author}
  {\bibfnamefont {H.~Q.}\ \bibnamefont {Xu}},\ }\bibfield  {title} {\bibinfo
  {title} {0-pi phase transition in hybrid superconductor--{InSb} nanowire
  quantum dot devices},\ }\href {https://doi.org/10.1103/PhysRevB.95.014515}
  {\bibfield  {journal} {\bibinfo  {journal} {Physical Review B}\ }\textbf
  {\bibinfo {volume} {95}},\ \bibinfo {pages} {014515} (\bibinfo {year}
  {2017})}\BibitemShut {NoStop}%
\bibitem [{\citenamefont {Kim}\ \emph {et~al.}(2013)\citenamefont {Kim},
  \citenamefont {Ahn}, \citenamefont {Kim}, \citenamefont {Choi}, \citenamefont
  {Bae}, \citenamefont {Kang}, \citenamefont {Lim}, \citenamefont {López},\
  and\ \citenamefont {Kim}}]{kim_transport_2013}%
  \BibitemOpen
  \bibfield  {author} {\bibinfo {author} {\bibfnamefont {B.-K.}\ \bibnamefont
  {Kim}}, \bibinfo {author} {\bibfnamefont {Y.-H.}\ \bibnamefont {Ahn}},
  \bibinfo {author} {\bibfnamefont {J.-J.}\ \bibnamefont {Kim}}, \bibinfo
  {author} {\bibfnamefont {M.-S.}\ \bibnamefont {Choi}}, \bibinfo {author}
  {\bibfnamefont {M.-H.}\ \bibnamefont {Bae}}, \bibinfo {author} {\bibfnamefont
  {K.}~\bibnamefont {Kang}}, \bibinfo {author} {\bibfnamefont {J.~S.}\
  \bibnamefont {Lim}}, \bibinfo {author} {\bibfnamefont {R.}~\bibnamefont
  {López}},\ and\ \bibinfo {author} {\bibfnamefont {N.}~\bibnamefont {Kim}},\
  }\bibfield  {title} {\bibinfo {title} {Transport {Measurement} of {Andreev}
  {Bound} {States} in a {Kondo}-{Correlated} {Quantum} {Dot}},\ }\href
  {https://doi.org/10.1103/PhysRevLett.110.076803} {\bibfield  {journal}
  {\bibinfo  {journal} {Physical Review Letters}\ }\textbf {\bibinfo {volume}
  {110}},\ \bibinfo {pages} {076803} (\bibinfo {year} {2013})}\BibitemShut
  {NoStop}%
\bibitem [{\citenamefont {Lee}\ \emph {et~al.}(2014)\citenamefont {Lee},
  \citenamefont {Jiang}, \citenamefont {Houzet}, \citenamefont {Aguado},
  \citenamefont {Lieber},\ and\ \citenamefont
  {De~Franceschi}}]{lee_spin-resolved_2014}%
  \BibitemOpen
  \bibfield  {author} {\bibinfo {author} {\bibfnamefont {E.~J.~H.}\
  \bibnamefont {Lee}}, \bibinfo {author} {\bibfnamefont {X.}~\bibnamefont
  {Jiang}}, \bibinfo {author} {\bibfnamefont {M.}~\bibnamefont {Houzet}},
  \bibinfo {author} {\bibfnamefont {R.}~\bibnamefont {Aguado}}, \bibinfo
  {author} {\bibfnamefont {C.~M.}\ \bibnamefont {Lieber}},\ and\ \bibinfo
  {author} {\bibfnamefont {S.}~\bibnamefont {De~Franceschi}},\ }\bibfield
  {title} {\bibinfo {title} {Spin-resolved {Andreev} levels and parity
  crossings in hybrid superconductor–semiconductor nanostructures},\ }\href
  {https://doi.org/10.1038/nnano.2013.267} {\bibfield  {journal} {\bibinfo
  {journal} {Nature Nanotechnology}\ }\textbf {\bibinfo {volume} {9}},\
  \bibinfo {pages} {79} (\bibinfo {year} {2014})}\BibitemShut {NoStop}%
\bibitem [{\citenamefont {Delagrange}\ \emph {et~al.}(2015)\citenamefont
  {Delagrange}, \citenamefont {Luitz}, \citenamefont {Weil}, \citenamefont
  {Kasumov}, \citenamefont {Meden}, \citenamefont {Bouchiat},\ and\
  \citenamefont {Deblock}}]{delagrange_manipulating_2015}%
  \BibitemOpen
  \bibfield  {author} {\bibinfo {author} {\bibfnamefont {R.}~\bibnamefont
  {Delagrange}}, \bibinfo {author} {\bibfnamefont {D.~J.}\ \bibnamefont
  {Luitz}}, \bibinfo {author} {\bibfnamefont {R.}~\bibnamefont {Weil}},
  \bibinfo {author} {\bibfnamefont {A.}~\bibnamefont {Kasumov}}, \bibinfo
  {author} {\bibfnamefont {V.}~\bibnamefont {Meden}}, \bibinfo {author}
  {\bibfnamefont {H.}~\bibnamefont {Bouchiat}},\ and\ \bibinfo {author}
  {\bibfnamefont {R.}~\bibnamefont {Deblock}},\ }\bibfield  {title} {\bibinfo
  {title} {Manipulating the magnetic state of a carbon nanotube {Josephson}
  junction using the superconducting phase},\ }\href
  {https://doi.org/10.1103/PhysRevB.91.241401} {\bibfield  {journal} {\bibinfo
  {journal} {Physical Review B}\ }\textbf {\bibinfo {volume} {91}},\ \bibinfo
  {pages} {241401} (\bibinfo {year} {2015})}\BibitemShut {NoStop}%
\bibitem [{\citenamefont {Szombati}\ \emph {et~al.}(2016)\citenamefont
  {Szombati}, \citenamefont {Nadj-Perge}, \citenamefont {Car}, \citenamefont
  {Plissard}, \citenamefont {Bakkers},\ and\ \citenamefont
  {Kouwenhoven}}]{szombati_josephson_2016}%
  \BibitemOpen
  \bibfield  {author} {\bibinfo {author} {\bibfnamefont {D.~B.}\ \bibnamefont
  {Szombati}}, \bibinfo {author} {\bibfnamefont {S.}~\bibnamefont
  {Nadj-Perge}}, \bibinfo {author} {\bibfnamefont {D.}~\bibnamefont {Car}},
  \bibinfo {author} {\bibfnamefont {S.~R.}\ \bibnamefont {Plissard}}, \bibinfo
  {author} {\bibfnamefont {E.~P. a.~M.}\ \bibnamefont {Bakkers}},\ and\
  \bibinfo {author} {\bibfnamefont {L.~P.}\ \bibnamefont {Kouwenhoven}},\
  }\bibfield  {title} {\bibinfo {title} {Josephson phi0-junction in nanowire
  quantum dots},\ }\href {https://doi.org/10.1038/nphys3742} {\bibfield
  {journal} {\bibinfo  {journal} {Nature Physics}\ }\textbf {\bibinfo {volume}
  {12}},\ \bibinfo {pages} {568} (\bibinfo {year} {2016})}\BibitemShut
  {NoStop}%
\bibitem [{\citenamefont {Lee}\ \emph {et~al.}(2017)\citenamefont {Lee},
  \citenamefont {Jiang}, \citenamefont {Žitko}, \citenamefont {Aguado},
  \citenamefont {Lieber},\ and\ \citenamefont
  {De~Franceschi}}]{lee_scaling_2017}%
  \BibitemOpen
  \bibfield  {author} {\bibinfo {author} {\bibfnamefont {E.~J.~H.}\
  \bibnamefont {Lee}}, \bibinfo {author} {\bibfnamefont {X.}~\bibnamefont
  {Jiang}}, \bibinfo {author} {\bibfnamefont {R.}~\bibnamefont {Žitko}},
  \bibinfo {author} {\bibfnamefont {R.}~\bibnamefont {Aguado}}, \bibinfo
  {author} {\bibfnamefont {C.~M.}\ \bibnamefont {Lieber}},\ and\ \bibinfo
  {author} {\bibfnamefont {S.}~\bibnamefont {De~Franceschi}},\ }\bibfield
  {title} {\bibinfo {title} {Scaling of subgap excitations in a
  superconductor-semiconductor nanowire quantum dot},\ }\href
  {https://doi.org/10.1103/PhysRevB.95.180502} {\bibfield  {journal} {\bibinfo
  {journal} {Physical Review B}\ }\textbf {\bibinfo {volume} {95}},\ \bibinfo
  {pages} {180502} (\bibinfo {year} {2017})}\BibitemShut {NoStop}%
\bibitem [{\citenamefont {Razmadze}\ \emph {et~al.}(2020)\citenamefont
  {Razmadze}, \citenamefont {O’Farrell}, \citenamefont {Krogstrup},\ and\
  \citenamefont {Marcus}}]{razmadze_quantum_2020}%
  \BibitemOpen
  \bibfield  {author} {\bibinfo {author} {\bibfnamefont {D.}~\bibnamefont
  {Razmadze}}, \bibinfo {author} {\bibfnamefont {E.}~\bibnamefont
  {O’Farrell}}, \bibinfo {author} {\bibfnamefont {P.}~\bibnamefont
  {Krogstrup}},\ and\ \bibinfo {author} {\bibfnamefont {C.}~\bibnamefont
  {Marcus}},\ }\bibfield  {title} {\bibinfo {title} {Quantum {Dot} {Parity}
  {Effects} in {Trivial} and {Topological} {Josephson} {Junctions}},\ }\href
  {https://doi.org/10.1103/PhysRevLett.125.116803} {\bibfield  {journal}
  {\bibinfo  {journal} {Physical Review Letters}\ }\textbf {\bibinfo {volume}
  {125}},\ \bibinfo {pages} {116803} (\bibinfo {year} {2020})}\BibitemShut
  {NoStop}%
\bibitem [{\citenamefont {Su}\ \emph {et~al.}(2020)\citenamefont {Su},
  \citenamefont {Žitko}, \citenamefont {Zhang}, \citenamefont {Wu},
  \citenamefont {Car}, \citenamefont {Plissard}, \citenamefont {Gazibegovic},
  \citenamefont {Badawy}, \citenamefont {Hocevar}, \citenamefont {Chen},
  \citenamefont {Bakkers},\ and\ \citenamefont {Frolov}}]{su_erasing_2020}%
  \BibitemOpen
  \bibfield  {author} {\bibinfo {author} {\bibfnamefont {Z.}~\bibnamefont
  {Su}}, \bibinfo {author} {\bibfnamefont {R.}~\bibnamefont {Žitko}}, \bibinfo
  {author} {\bibfnamefont {P.}~\bibnamefont {Zhang}}, \bibinfo {author}
  {\bibfnamefont {H.}~\bibnamefont {Wu}}, \bibinfo {author} {\bibfnamefont
  {D.}~\bibnamefont {Car}}, \bibinfo {author} {\bibfnamefont {S.~R.}\
  \bibnamefont {Plissard}}, \bibinfo {author} {\bibfnamefont {S.}~\bibnamefont
  {Gazibegovic}}, \bibinfo {author} {\bibfnamefont {G.}~\bibnamefont {Badawy}},
  \bibinfo {author} {\bibfnamefont {M.}~\bibnamefont {Hocevar}}, \bibinfo
  {author} {\bibfnamefont {J.}~\bibnamefont {Chen}}, \bibinfo {author}
  {\bibfnamefont {E.~P. A.~M.}\ \bibnamefont {Bakkers}},\ and\ \bibinfo
  {author} {\bibfnamefont {S.~M.}\ \bibnamefont {Frolov}},\ }\bibfield  {title}
  {\bibinfo {title} {Erasing odd-parity states in semiconductor quantum dots
  coupled to superconductors},\ }\href
  {https://doi.org/10.1103/PhysRevB.101.235315} {\bibfield  {journal} {\bibinfo
   {journal} {Physical Review B}\ }\textbf {\bibinfo {volume} {101}},\ \bibinfo
  {pages} {235315} (\bibinfo {year} {2020})}\BibitemShut {NoStop}%
\bibitem [{\citenamefont {Kulik}(1970)}]{kulik_macroscopic_1970}%
  \BibitemOpen
  \bibfield  {author} {\bibinfo {author} {\bibfnamefont {I.}~\bibnamefont
  {Kulik}},\ }\bibfield  {title} {\bibinfo {title} {Macroscopic {Quantization}
  and the {Proximity} {Effect} in {S}-{N}-{S} {Junctions}},\ }\href
  {http://jetp.ac.ru/cgi-bin/e/index/e/30/5/p944?a=list} {\bibfield  {journal}
  {\bibinfo  {journal} {JETP}\ }\textbf {\bibinfo {volume} {30}},\ \bibinfo
  {pages} {944} (\bibinfo {year} {1970})}\BibitemShut {NoStop}%
\bibitem [{\citenamefont {Beenakker}\ and\ \citenamefont
  {Houten}(1992)}]{beenakker_resonant_1992}%
  \BibitemOpen
  \bibfield  {author} {\bibinfo {author} {\bibfnamefont {C.~W.~J.}\
  \bibnamefont {Beenakker}}\ and\ \bibinfo {author} {\bibfnamefont {H.~v.}\
  \bibnamefont {Houten}},\ }\bibfield  {title} {\bibinfo {title} {Resonant
  {Josephson} {Current} {Through} a {Quantum} {Dot}},\ }in\ \href
  {https://link.springer.com/chapter/10.1007/978-3-642-77274-0_20} {\emph
  {\bibinfo {booktitle} {Single-{Electron} {Tunneling} and {Mesoscopic}
  {Devices}}}},\ \bibinfo {series and number} {Springer {Series} in
  {Electronics} and {Photonics}}\ (\bibinfo  {publisher} {Springer, Berlin,
  Heidelberg},\ \bibinfo {year} {1992})\ pp.\ \bibinfo {pages}
  {175--179}\BibitemShut {NoStop}%
\bibitem [{\citenamefont {Meden}(2019)}]{meden_andersonjosephson_2019}%
  \BibitemOpen
  \bibfield  {author} {\bibinfo {author} {\bibfnamefont {V.}~\bibnamefont
  {Meden}},\ }\bibfield  {title} {\bibinfo {title} {The
  {Anderson}–{Josephson} quantum dot—a theory perspective},\ }\href
  {https://doi.org/10.1088/1361-648X/aafd6a} {\bibfield  {journal} {\bibinfo
  {journal} {Journal of Physics: Condensed Matter}\ }\textbf {\bibinfo {volume}
  {31}},\ \bibinfo {pages} {163001} (\bibinfo {year} {2019})}\BibitemShut
  {NoStop}%
\bibitem [{\citenamefont {Blais}\ \emph {et~al.}(2021)\citenamefont {Blais},
  \citenamefont {Grimsmo}, \citenamefont {Girvin},\ and\ \citenamefont
  {Wallraff}}]{blais_circuit_2021}%
  \BibitemOpen
  \bibfield  {author} {\bibinfo {author} {\bibfnamefont {A.}~\bibnamefont
  {Blais}}, \bibinfo {author} {\bibfnamefont {A.~L.}\ \bibnamefont {Grimsmo}},
  \bibinfo {author} {\bibfnamefont {S.}~\bibnamefont {Girvin}},\ and\ \bibinfo
  {author} {\bibfnamefont {A.}~\bibnamefont {Wallraff}},\ }\bibfield  {title}
  {\bibinfo {title} {Circuit quantum electrodynamics},\ }\href
  {https://doi.org/10.1103/RevModPhys.93.025005} {\bibfield  {journal}
  {\bibinfo  {journal} {Reviews of Modern Physics}\ }\textbf {\bibinfo {volume}
  {93}},\ \bibinfo {pages} {025005} (\bibinfo {year} {2021})},\ \bibinfo {note}
  {tex.ids= blais\_circuit\_2020 arXiv: 2005.12667 publisher: American Physical
  Society}\BibitemShut {NoStop}%
\bibitem [{\citenamefont {Paila}\ \emph {et~al.}(2009)\citenamefont {Paila},
  \citenamefont {Gunnarsson}, \citenamefont {Sarkar}, \citenamefont
  {Sillanpää},\ and\ \citenamefont {Hakonen}}]{paila_current-phase_2009}%
  \BibitemOpen
  \bibfield  {author} {\bibinfo {author} {\bibfnamefont {A.}~\bibnamefont
  {Paila}}, \bibinfo {author} {\bibfnamefont {D.}~\bibnamefont {Gunnarsson}},
  \bibinfo {author} {\bibfnamefont {J.}~\bibnamefont {Sarkar}}, \bibinfo
  {author} {\bibfnamefont {M.~A.}\ \bibnamefont {Sillanpää}},\ and\ \bibinfo
  {author} {\bibfnamefont {P.~J.}\ \bibnamefont {Hakonen}},\ }\bibfield
  {title} {\bibinfo {title} {Current-phase relation and {Josephson} inductance
  in a superconducting {Cooper}-pair transistor},\ }\href
  {https://doi.org/10.1103/PhysRevB.80.144520} {\bibfield  {journal} {\bibinfo
  {journal} {Physical Review B}\ }\textbf {\bibinfo {volume} {80}},\ \bibinfo
  {pages} {144520} (\bibinfo {year} {2009})}\BibitemShut {NoStop}%
\bibitem [{\citenamefont {Basov}\ \emph {et~al.}(2011)\citenamefont {Basov},
  \citenamefont {Averitt}, \citenamefont {van~der Marel}, \citenamefont
  {Dressel},\ and\ \citenamefont {Haule}}]{basov_electrodynamics_2011}%
  \BibitemOpen
  \bibfield  {author} {\bibinfo {author} {\bibfnamefont {D.~N.}\ \bibnamefont
  {Basov}}, \bibinfo {author} {\bibfnamefont {R.~D.}\ \bibnamefont {Averitt}},
  \bibinfo {author} {\bibfnamefont {D.}~\bibnamefont {van~der Marel}}, \bibinfo
  {author} {\bibfnamefont {M.}~\bibnamefont {Dressel}},\ and\ \bibinfo {author}
  {\bibfnamefont {K.}~\bibnamefont {Haule}},\ }\bibfield  {title} {\bibinfo
  {title} {Electrodynamics of correlated electron materials},\ }\href
  {https://doi.org/10.1103/RevModPhys.83.471} {\bibfield  {journal} {\bibinfo
  {journal} {Reviews of Modern Physics}\ }\textbf {\bibinfo {volume} {83}},\
  \bibinfo {pages} {471} (\bibinfo {year} {2011})}\BibitemShut {NoStop}%
\bibitem [{\citenamefont {Metzger}\ \emph {et~al.}(2021)\citenamefont
  {Metzger}, \citenamefont {Park}, \citenamefont {Tosi}, \citenamefont
  {Janvier}, \citenamefont {Reynoso}, \citenamefont {Goffman}, \citenamefont
  {Urbina}, \citenamefont {Levy~Yeyati},\ and\ \citenamefont
  {Pothier}}]{metzger_circuit-qed_2021}%
  \BibitemOpen
  \bibfield  {author} {\bibinfo {author} {\bibfnamefont {C.}~\bibnamefont
  {Metzger}}, \bibinfo {author} {\bibfnamefont {S.}~\bibnamefont {Park}},
  \bibinfo {author} {\bibfnamefont {L.}~\bibnamefont {Tosi}}, \bibinfo {author}
  {\bibfnamefont {C.}~\bibnamefont {Janvier}}, \bibinfo {author} {\bibfnamefont
  {A.~A.}\ \bibnamefont {Reynoso}}, \bibinfo {author} {\bibfnamefont {M.~F.}\
  \bibnamefont {Goffman}}, \bibinfo {author} {\bibfnamefont {C.}~\bibnamefont
  {Urbina}}, \bibinfo {author} {\bibfnamefont {A.}~\bibnamefont
  {Levy~Yeyati}},\ and\ \bibinfo {author} {\bibfnamefont {H.}~\bibnamefont
  {Pothier}},\ }\bibfield  {title} {\bibinfo {title} {Circuit-{QED} with
  phase-biased {Josephson} weak links},\ }\href
  {https://doi.org/10.1103/PhysRevResearch.3.013036} {\bibfield  {journal}
  {\bibinfo  {journal} {Physical Review Research}\ }\textbf {\bibinfo {volume}
  {3}},\ \bibinfo {pages} {013036} (\bibinfo {year} {2021})}\BibitemShut
  {NoStop}%
\bibitem [{\citenamefont {Haller}\ \emph {et~al.}(2021)\citenamefont {Haller},
  \citenamefont {Fülöp}, \citenamefont {Indolese}, \citenamefont {Ridderbos},
  \citenamefont {Kraft}, \citenamefont {Cheung}, \citenamefont {Ungerer},
  \citenamefont {Watanabe}, \citenamefont {Taniguchi}, \citenamefont
  {Beckmann}, \citenamefont {Danneau}, \citenamefont {Virtanen},\ and\
  \citenamefont {Schönenberger}}]{haller_phase-dependent_2021}%
  \BibitemOpen
  \bibfield  {author} {\bibinfo {author} {\bibfnamefont {R.}~\bibnamefont
  {Haller}}, \bibinfo {author} {\bibfnamefont {G.}~\bibnamefont {Fülöp}},
  \bibinfo {author} {\bibfnamefont {D.}~\bibnamefont {Indolese}}, \bibinfo
  {author} {\bibfnamefont {J.}~\bibnamefont {Ridderbos}}, \bibinfo {author}
  {\bibfnamefont {R.}~\bibnamefont {Kraft}}, \bibinfo {author} {\bibfnamefont
  {L.~Y.}\ \bibnamefont {Cheung}}, \bibinfo {author} {\bibfnamefont {J.~H.}\
  \bibnamefont {Ungerer}}, \bibinfo {author} {\bibfnamefont {K.}~\bibnamefont
  {Watanabe}}, \bibinfo {author} {\bibfnamefont {T.}~\bibnamefont {Taniguchi}},
  \bibinfo {author} {\bibfnamefont {D.}~\bibnamefont {Beckmann}}, \bibinfo
  {author} {\bibfnamefont {R.}~\bibnamefont {Danneau}}, \bibinfo {author}
  {\bibfnamefont {P.}~\bibnamefont {Virtanen}},\ and\ \bibinfo {author}
  {\bibfnamefont {C.}~\bibnamefont {Schönenberger}},\ }\bibfield  {title}
  {\bibinfo {title} {Phase-dependent microwave response of a graphene
  {Josephson} junction},\ }\href {http://arxiv.org/abs/2108.00989} {\bibfield
  {journal} {\bibinfo  {journal} {arXiv:2108.00989 [cond-mat]}\ } (\bibinfo
  {year} {2021})},\ \bibinfo {note} {arXiv: 2108.00989}\BibitemShut {NoStop}%
\bibitem [{\citenamefont {Ivanov}\ and\ \citenamefont
  {Feigel’man}(1999)}]{ivanov_two-level_1999}%
  \BibitemOpen
  \bibfield  {author} {\bibinfo {author} {\bibfnamefont {D.~A.}\ \bibnamefont
  {Ivanov}}\ and\ \bibinfo {author} {\bibfnamefont {M.~V.}\ \bibnamefont
  {Feigel’man}},\ }\bibfield  {title} {\bibinfo {title} {Two-level
  {Hamiltonian} of a superconducting quantum point contact},\ }\href
  {https://doi.org/10.1103/PhysRevB.59.8444} {\bibfield  {journal} {\bibinfo
  {journal} {Physical Review B}\ }\textbf {\bibinfo {volume} {59}},\ \bibinfo
  {pages} {8444} (\bibinfo {year} {1999})}\BibitemShut {NoStop}%
\bibitem [{\citenamefont {Zazunov}\ \emph {et~al.}(2003)\citenamefont
  {Zazunov}, \citenamefont {Shumeiko}, \citenamefont {Bratus’}, \citenamefont
  {Lantz},\ and\ \citenamefont {Wendin}}]{zazunov_andreev_2003}%
  \BibitemOpen
  \bibfield  {author} {\bibinfo {author} {\bibfnamefont {A.}~\bibnamefont
  {Zazunov}}, \bibinfo {author} {\bibfnamefont {V.~S.}\ \bibnamefont
  {Shumeiko}}, \bibinfo {author} {\bibfnamefont {E.~N.}\ \bibnamefont
  {Bratus’}}, \bibinfo {author} {\bibfnamefont {J.}~\bibnamefont {Lantz}},\
  and\ \bibinfo {author} {\bibfnamefont {G.}~\bibnamefont {Wendin}},\
  }\bibfield  {title} {\bibinfo {title} {Andreev {Level} {Qubit}},\ }\href
  {https://doi.org/10.1103/PhysRevLett.90.087003} {\bibfield  {journal}
  {\bibinfo  {journal} {Physical Review Letters}\ }\textbf {\bibinfo {volume}
  {90}},\ \bibinfo {pages} {087003} (\bibinfo {year} {2003})}\BibitemShut
  {NoStop}%
\bibitem [{\citenamefont {Zazunov}\ \emph {et~al.}(2005)\citenamefont
  {Zazunov}, \citenamefont {Shumeiko}, \citenamefont {Wendin},\ and\
  \citenamefont {Bratus’}}]{zazunov_dynamics_2005}%
  \BibitemOpen
  \bibfield  {author} {\bibinfo {author} {\bibfnamefont {A.}~\bibnamefont
  {Zazunov}}, \bibinfo {author} {\bibfnamefont {V.~S.}\ \bibnamefont
  {Shumeiko}}, \bibinfo {author} {\bibfnamefont {G.}~\bibnamefont {Wendin}},\
  and\ \bibinfo {author} {\bibfnamefont {E.~N.}\ \bibnamefont {Bratus’}},\
  }\bibfield  {title} {\bibinfo {title} {Dynamics and phonon-induced
  decoherence of {Andreev} level qubit},\ }\href
  {https://doi.org/10.1103/PhysRevB.71.214505} {\bibfield  {journal} {\bibinfo
  {journal} {Physical Review B}\ }\textbf {\bibinfo {volume} {71}},\ \bibinfo
  {pages} {214505} (\bibinfo {year} {2005})}\BibitemShut {NoStop}%
\bibitem [{\citenamefont {Wendin}\ and\ \citenamefont
  {Shumeiko}(1996)}]{wendin_josephson_1996}%
  \BibitemOpen
  \bibfield  {author} {\bibinfo {author} {\bibfnamefont {G.}~\bibnamefont
  {Wendin}}\ and\ \bibinfo {author} {\bibfnamefont {V.~S.}\ \bibnamefont
  {Shumeiko}},\ }\bibfield  {title} {\bibinfo {title} {Josephson transport in
  complex mesoscopic structures},\ }\href
  {https://doi.org/10.1006/spmi.1996.0116} {\bibfield  {journal} {\bibinfo
  {journal} {Superlattices and Microstructures}\ }\textbf {\bibinfo {volume}
  {20}},\ \bibinfo {pages} {569} (\bibinfo {year} {1996})}\BibitemShut
  {NoStop}%
\bibitem [{\citenamefont {Kurilovich}\ \emph {et~al.}(2021)\citenamefont
  {Kurilovich}, \citenamefont {Kurilovich}, \citenamefont {Fatemi},
  \citenamefont {Devoret},\ and\ \citenamefont
  {Glazman}}]{kurilovich_microwave_2021}%
  \BibitemOpen
  \bibfield  {author} {\bibinfo {author} {\bibfnamefont {P.~D.}\ \bibnamefont
  {Kurilovich}}, \bibinfo {author} {\bibfnamefont {V.~D.}\ \bibnamefont
  {Kurilovich}}, \bibinfo {author} {\bibfnamefont {V.}~\bibnamefont {Fatemi}},
  \bibinfo {author} {\bibfnamefont {M.~H.}\ \bibnamefont {Devoret}},\ and\
  \bibinfo {author} {\bibfnamefont {L.~I.}\ \bibnamefont {Glazman}},\
  }\bibfield  {title} {\bibinfo {title} {Microwave response of an {Andreev}
  bound state},\ }\href {https://doi.org/10.1103/PhysRevB.104.174517}
  {\bibfield  {journal} {\bibinfo  {journal} {Physical Review B}\ }\textbf
  {\bibinfo {volume} {104}},\ \bibinfo {pages} {174517} (\bibinfo {year}
  {2021})}\BibitemShut {NoStop}%
\bibitem [{\citenamefont {van Woerkom}\ \emph {et~al.}(2018)\citenamefont {van
  Woerkom}, \citenamefont {Scarlino}, \citenamefont {Ungerer}, \citenamefont
  {Müller}, \citenamefont {Koski}, \citenamefont {Landig}, \citenamefont
  {Reichl}, \citenamefont {Wegscheider}, \citenamefont {Ihn}, \citenamefont
  {Ensslin},\ and\ \citenamefont {Wallraff}}]{van_woerkom_microwave_2018}%
  \BibitemOpen
  \bibfield  {author} {\bibinfo {author} {\bibfnamefont {D.}~\bibnamefont {van
  Woerkom}}, \bibinfo {author} {\bibfnamefont {P.}~\bibnamefont {Scarlino}},
  \bibinfo {author} {\bibfnamefont {J.}~\bibnamefont {Ungerer}}, \bibinfo
  {author} {\bibfnamefont {C.}~\bibnamefont {Müller}}, \bibinfo {author}
  {\bibfnamefont {J.}~\bibnamefont {Koski}}, \bibinfo {author} {\bibfnamefont
  {A.}~\bibnamefont {Landig}}, \bibinfo {author} {\bibfnamefont
  {C.}~\bibnamefont {Reichl}}, \bibinfo {author} {\bibfnamefont
  {W.}~\bibnamefont {Wegscheider}}, \bibinfo {author} {\bibfnamefont
  {T.}~\bibnamefont {Ihn}}, \bibinfo {author} {\bibfnamefont {K.}~\bibnamefont
  {Ensslin}},\ and\ \bibinfo {author} {\bibfnamefont {A.}~\bibnamefont
  {Wallraff}},\ }\bibfield  {title} {\bibinfo {title} {Microwave
  {Photon}-{Mediated} {Interactions} between {Semiconductor} {Qubits}},\ }\href
  {https://doi.org/10.1103/PhysRevX.8.041018} {\bibfield  {journal} {\bibinfo
  {journal} {Physical Review X}\ }\textbf {\bibinfo {volume} {8}},\ \bibinfo
  {pages} {041018} (\bibinfo {year} {2018})}\BibitemShut {NoStop}%
\bibitem [{Note1()}]{Note1}%
  \BibitemOpen
  \bibinfo {note} {Models with weak link length of order the coherence length
  also exhibit dwell time. Such models admit additional doublets and add
  theoretical complications, see supplement V.E.}\BibitemShut {Stop}%
\bibitem [{Note2()}]{Note2}%
  \BibitemOpen
  \bibinfo {note} {Probe tone power corresponding to roughly 9 photons in the
  resonator.}\BibitemShut {Stop}%
\bibitem [{Note3()}]{Note3}%
  \BibitemOpen
  \bibinfo {note} {This na{\"\i }ve comparison is in principle susceptible to
  corrections to the dispersive shifts because they are solutions of a
  transcendental equation which is inherently nonlinear in $Y(\omega )$ (see
  supplement). We estimate the corrections to be less than 10\% of the mismatch
  observed here. Furthermore, our probe power is small enough that frequency
  shifts due to nonlinearities are minimal.}\BibitemShut {Stop}%
\bibitem [{\citenamefont {Bargerbos}\ \emph {et~al.}(2020)\citenamefont
  {Bargerbos}, \citenamefont {Uilhoorn}, \citenamefont {Yang}, \citenamefont
  {Krogstrup}, \citenamefont {Kouwenhoven}, \citenamefont {de~Lange},
  \citenamefont {van Heck},\ and\ \citenamefont
  {Kou}}]{bargerbos_observation_2020}%
  \BibitemOpen
  \bibfield  {author} {\bibinfo {author} {\bibfnamefont {A.}~\bibnamefont
  {Bargerbos}}, \bibinfo {author} {\bibfnamefont {W.}~\bibnamefont {Uilhoorn}},
  \bibinfo {author} {\bibfnamefont {C.-K.}\ \bibnamefont {Yang}}, \bibinfo
  {author} {\bibfnamefont {P.}~\bibnamefont {Krogstrup}}, \bibinfo {author}
  {\bibfnamefont {L.~P.}\ \bibnamefont {Kouwenhoven}}, \bibinfo {author}
  {\bibfnamefont {G.}~\bibnamefont {de~Lange}}, \bibinfo {author}
  {\bibfnamefont {B.}~\bibnamefont {van Heck}},\ and\ \bibinfo {author}
  {\bibfnamefont {A.}~\bibnamefont {Kou}},\ }\bibfield  {title} {\bibinfo
  {title} {Observation of {Vanishing} {Charge} {Dispersion} of a {Nearly}
  {Open} {Superconducting} {Island}},\ }\href
  {https://doi.org/10.1103/PhysRevLett.124.246802} {\bibfield  {journal}
  {\bibinfo  {journal} {Physical Review Letters}\ }\textbf {\bibinfo {volume}
  {124}},\ \bibinfo {pages} {246802} (\bibinfo {year} {2020})}\BibitemShut
  {NoStop}%
\bibitem [{\citenamefont {Kringhøj}\ \emph {et~al.}(2020)\citenamefont
  {Kringhøj}, \citenamefont {van Heck}, \citenamefont {Larsen}, \citenamefont
  {Erlandsson}, \citenamefont {Sabonis}, \citenamefont {Krogstrup},
  \citenamefont {Casparis}, \citenamefont {Petersson},\ and\ \citenamefont
  {Marcus}}]{kringhoj_suppressed_2020}%
  \BibitemOpen
  \bibfield  {author} {\bibinfo {author} {\bibfnamefont {A.}~\bibnamefont
  {Kringhøj}}, \bibinfo {author} {\bibfnamefont {B.}~\bibnamefont {van Heck}},
  \bibinfo {author} {\bibfnamefont {T.}~\bibnamefont {Larsen}}, \bibinfo
  {author} {\bibfnamefont {O.}~\bibnamefont {Erlandsson}}, \bibinfo {author}
  {\bibfnamefont {D.}~\bibnamefont {Sabonis}}, \bibinfo {author} {\bibfnamefont
  {P.}~\bibnamefont {Krogstrup}}, \bibinfo {author} {\bibfnamefont
  {L.}~\bibnamefont {Casparis}}, \bibinfo {author} {\bibfnamefont
  {K.}~\bibnamefont {Petersson}},\ and\ \bibinfo {author} {\bibfnamefont
  {C.}~\bibnamefont {Marcus}},\ }\bibfield  {title} {\bibinfo {title}
  {Suppressed {Charge} {Dispersion} via {Resonant} {Tunneling} in a
  {Single}-{Channel} {Transmon}},\ }\href
  {https://doi.org/10.1103/PhysRevLett.124.246803} {\bibfield  {journal}
  {\bibinfo  {journal} {Physical Review Letters}\ }\textbf {\bibinfo {volume}
  {124}},\ \bibinfo {pages} {246803} (\bibinfo {year} {2020})}\BibitemShut
  {NoStop}%
\bibitem [{Note4()}]{Note4}%
  \BibitemOpen
  \bibinfo {note} {We note that a naive estimation based on the dimensions of
  just the uncovered weak link region gives of order $U \sim \SI {200}{\giga
  \hertz }$, whereas our fits produce values of $U$ that are up to an order of
  magnitude smaller. This suggests strong screening by the epitaxial aluminum
  leads.}\BibitemShut {Stop}%
\bibitem [{\citenamefont {Glazman}\ and\ \citenamefont
  {Catelani}(2021)}]{glazman_bogoliubov_2021}%
  \BibitemOpen
  \bibfield  {author} {\bibinfo {author} {\bibfnamefont {L.}~\bibnamefont
  {Glazman}}\ and\ \bibinfo {author} {\bibfnamefont {G.}~\bibnamefont
  {Catelani}},\ }\bibfield  {title} {\bibinfo {title} {Bogoliubov
  quasiparticles in superconducting qubits},\ }\bibfield  {journal} {\bibinfo
  {journal} {SciPost Physics Lecture Notes}\ }\href
  {https://doi.org/10.21468/SciPostPhysLectNotes.31}
  {10.21468/SciPostPhysLectNotes.31} (\bibinfo {year} {2021})\BibitemShut
  {NoStop}%
\bibitem [{\citenamefont {Matute~Canadas}\ \emph {et~al.}(2021)\citenamefont
  {Matute~Canadas}, \citenamefont {Metzger}, \citenamefont {Park},
  \citenamefont {Tosi}, \citenamefont {Krogstrup}, \citenamefont {Nygard},
  \citenamefont {Goffman}, \citenamefont {Urbina}, \citenamefont {Pothier},\
  and\ \citenamefont {Levy~Yeyati}}]{matute_canadas_signatures_2021}%
  \BibitemOpen
  \bibfield  {author} {\bibinfo {author} {\bibfnamefont {F.~J.}\ \bibnamefont
  {Matute~Canadas}}, \bibinfo {author} {\bibfnamefont {C.}~\bibnamefont
  {Metzger}}, \bibinfo {author} {\bibfnamefont {S.}~\bibnamefont {Park}},
  \bibinfo {author} {\bibfnamefont {L.}~\bibnamefont {Tosi}}, \bibinfo {author}
  {\bibfnamefont {P.}~\bibnamefont {Krogstrup}}, \bibinfo {author}
  {\bibfnamefont {J.}~\bibnamefont {Nygard}}, \bibinfo {author} {\bibfnamefont
  {M.~F.}\ \bibnamefont {Goffman}}, \bibinfo {author} {\bibfnamefont
  {C.}~\bibnamefont {Urbina}}, \bibinfo {author} {\bibfnamefont
  {H.}~\bibnamefont {Pothier}},\ and\ \bibinfo {author} {\bibfnamefont
  {A.}~\bibnamefont {Levy~Yeyati}},\ }\bibfield  {title} {\bibinfo {title}
  {Signatures of interactions in the {Andreev} spectrum of nanowire {Josephson}
  junctions},\ }\href@noop {} {\bibfield  {journal} {\bibinfo  {journal} {In
  Preparation}\ } (\bibinfo {year} {2021})}\BibitemShut {NoStop}%
\bibitem [{\citenamefont {Bargerbos}\ \emph {et~al.}()\citenamefont
  {Bargerbos}, \citenamefont {Pita-Vidal}, \citenamefont {Zitko}, \citenamefont
  {Avila}, \citenamefont {Splitthoff}, \citenamefont {Grunhaupt}, \citenamefont
  {Wesdorp}, \citenamefont {Andersen}, \citenamefont {Liu}, \citenamefont
  {Krogstrup}, \citenamefont {Kouwenhoven}, \citenamefont {Aguado},
  \citenamefont {Kou},\ and\ \citenamefont {van
  Heck}}]{bargerbos_singlet-doublet_nodate}%
  \BibitemOpen
  \bibfield  {author} {\bibinfo {author} {\bibfnamefont {A.}~\bibnamefont
  {Bargerbos}}, \bibinfo {author} {\bibfnamefont {M.}~\bibnamefont
  {Pita-Vidal}}, \bibinfo {author} {\bibfnamefont {R.}~\bibnamefont {Zitko}},
  \bibinfo {author} {\bibfnamefont {J.}~\bibnamefont {Avila}}, \bibinfo
  {author} {\bibfnamefont {L.~J.}\ \bibnamefont {Splitthoff}}, \bibinfo
  {author} {\bibfnamefont {L.}~\bibnamefont {Grunhaupt}}, \bibinfo {author}
  {\bibfnamefont {J.~J.}\ \bibnamefont {Wesdorp}}, \bibinfo {author}
  {\bibfnamefont {C.~K.}\ \bibnamefont {Andersen}}, \bibinfo {author}
  {\bibfnamefont {Y.}~\bibnamefont {Liu}}, \bibinfo {author} {\bibfnamefont
  {P.}~\bibnamefont {Krogstrup}}, \bibinfo {author} {\bibfnamefont {L.~P.}\
  \bibnamefont {Kouwenhoven}}, \bibinfo {author} {\bibfnamefont
  {R.}~\bibnamefont {Aguado}}, \bibinfo {author} {\bibfnamefont
  {A.}~\bibnamefont {Kou}},\ and\ \bibinfo {author} {\bibfnamefont
  {B.}~\bibnamefont {van Heck}},\ }\bibfield  {title} {\bibinfo {title}
  {Singlet-doublet transitions of a quantum dot {Josephson} junction revealed
  in a transmon circuit},\ }\href@noop {} {\bibinfo  {journal} {In
  Preparation}\ }\BibitemShut {NoStop}%
\end{thebibliography}%

\end{document}